\documentclass[AMA,STIX1COL]{WileyNJD-v2}
\usepackage{moreverb}

\newcommand\BibTeX{{\rmfamily B\kern-.05em \textsc{i\kern-.025em b}\kern-.08em
T\kern-.1667em\lower.7ex\hbox{E}\kern-.125emX}}

\articletype{Article Type}%

\usepackage{graphicx}
\usepackage{caption}
\usepackage{amsfonts}
\usepackage{amsmath,bm,bbm,amsthm}
\usepackage{titlesec}
\usepackage{indentfirst}
\usepackage{times}
\usepackage{fancyhdr}
\usepackage{soul}
\usepackage[utf8]{inputenc}
\usepackage{amsmath}

\usepackage[algo2e,ruled,vlined]{algorithm2e} 

\usepackage{booktabs}
\usepackage{verbatim}
\usepackage{color}
\usepackage{tcolorbox}
\usepackage{mathtools}
\usepackage{hyperref}
\usepackage{footnote}
\usepackage[bottom]{footmisc}

\begin{document}

\title{A resampling approach for causal inference on novel two-point time-series with application to 
identify risk factors for type-2 diabetes and cardiovascular disease}

\author[1]{Xiaowu Dai*}

\author[2]{Saad Mouti*}

\author[3]{Marjorie Lima do Vale}

\author[3,4]{Sumantra Ray}

\author[5]{Jeffrey Bohn}

\author[6]{Lisa Goldberg}


\address[1]{\orgdiv{Department of Statistics}, \orgname{University of California, Los Angeles}, \orgaddress{\state{CA}, \country{USA}}}

\address[2]{\orgdiv{Statistics and Applied Probability}, \orgname{University of California, Santa Barbara}, \orgaddress{\state{CA}, \country{USA}}}

\address[3]{\orgname{NNEdPro Global Centre for Nutrition and Health}, \orgaddress{\state{Cambridge}, \country{UK}}}

\address[4]{\orgdiv{School of Biomedical Sciences},\orgname{University Of Ulster}; \orgdiv{School of the Humanities and Social Sciences},\orgname{University of Cambridge}, \orgaddress{\state{Cambridge}, \country{UK}}}

\address[5]{\orgdiv{CDAR}, \orgname{University of California, Berkeley}, \orgaddress{\state{CA}, \country{USA}}}

\address[6]{\orgdiv{Department of Economics and CDAR}, \orgname{University of California, Berkeley}, \orgaddress{\state{CA}, \country{USA}}}

\corres{Xiaowu Dai, \orgdiv{Department of Statistics}, \orgname{University of California, Los Angeles}.\\ \email{dai@stat.ucla.edu}}


\abstract[Abstract]{
Two-point time-series data, characterized by baseline and follow-up observations,  are frequently encountered in health research. 
We study a novel  two-point time-series structure without a control group, which is driven by an observational routine clinical dataset collected to monitor key risk markers of type-$2$ diabetes (T2D) and cardiovascular disease (CVD). 
We propose a resampling approach called `I-Rand' for independently sampling one of the two time points for each individual and making inference on the estimated causal effects based on matching methods.
The proposed method is illustrated with data from a service-based dietary intervention to promote a low-carbohydrate diet (LCD), designed to impact risk of T2D and CVD. 
Baseline data contain a pre-intervention health record of study participants, and health data after LCD intervention are recorded at the follow-up visit, providing a two-point time-series pattern without a parallel control group. Using this approach we find that obesity is a significant risk factor of T2D and CVD, and an LCD approach can significantly mitigate the risks of T2D and CVD. We provide code that implements our method.}


\maketitle

\section{Introduction}\label{sec1}

Cardiovascular disease (CVD) including stroke and  coronary heart diseases, 
has become the most common non-communicable disease in the United States, and is also a severe problem globally 
\cite{roth2018burden, unwin2020insights}. 
Type-2 diabetes (T2D) doubles the risk of CVD, which is the principal cause of death in T2D patients \cite{morrish2001mortality}.
CVD and T2D produce an immense economic burden on health care systems globally. 
Targeted intervention for individuals at increased risk of CVD and T2D plays a crucial role in reducing the global burden of these diseases\cite{Jan2018}. Consequently,
the identification of dietary and lifestyle risk factors for T2D and CVD has become a health priority \cite{benjamin2018}.  Since obesity is a substantial contributor to T2D, and consequently to the risk of CVD, \cite{Scheen2014} lowering obesity through diet control may help to alleviate the T2D and CVD epidemics.

In this work, we pursue two scientific goals.  First, we seek to determine whether or not obesity is a significant risk factor for T2D and CVD.  Second we ask if a low-carbohydrate diet (LCD) improves on standard care for T2D and CVD risk in patients with prediabetes or diabetes. 
We use causal inference tools, including the potential outcome model and mediation analysis, to quantify the impact of obesity and diet on T2D and CVD risk. To explore the link between obesity and T2D, we ask: \emph{what would the effect on T2D be if an individual were to change from a normal weight to an obese weight?}
Motivated by the impact of T2D in  CVD risk, we seek to understand the role of T2D in mediating the effect of obesity on CVD risk. 
This mediation analysis is relevant to 
an individual with limited control over his or her T2D status and who wishes to identify factors that can be controlled.
We perform mediation analysis to identify obesity as a significant risk factor for T2D and CVD and to disentangle cause-and-effect relationships in individuals with both conditions. 
Building on these questions, we are also interested in quantifying
the effects of an LCD,
which restricts the consumption of carbohydrates relative to the average diet\cite{Bazzano2014}, on both T2D and CVD risk. 
Several systematic reviews and meta-analyses of randomized control trials suggest beneficial effects of LCD in T2D and CVD \cite{meng2017efficacy, gjuladin2019effects, van2018effects}. However, the impact of LCD in a primary care setting with observational data and its cause-and-effect inferences has not been thoroughly evaluated \cite{unwin2020insights, lean2018primary,do2021synthesis}. 
As we discuss in detail later in this article, our results indicate that obesity is a significant risk factor for T2D and CVD, and that LCD can significantly lower the risks of T2D and CVD risk.

We explore our scientific questions by analyzing clinical data from patients who visited a health clinic in the UK on two occasions.  \color{black}{These patients began a low-carbohydrate diet subsequent to the first visit}, and standard measurements of their health were taken at both visits. Data on these patients naturally comprise a panel dataset with two time points.  In this two-point time-series dataset, there is no control group, which poses a challenge for causal inference.
We propose a novel approach to dealing with this challenge,  ``I-Rand," which estimates average treatment effect and its significance on a  collection of sub-samples of our dataset.  Each subsample contains exactly one of the two observations corresponding to each individual. The average treatment effect within each subsample relies on propensity score matching, and statistical significance is estimated with a permutation test.
Such subsampling has  been used  previously by Hahn \cite{Hahn2012} in the analysis of spatial point patterns.
We benchmark I-Rand against two alternative estimation methods. 
The I-Rand algorithm meets the Stable Unit Treatment Value Assumption (SUTVA) of  ``no-interference'' for valid causal inference, unlike the pooled approach\cite{beck1995and, wilson2007lot}. On the other hand, I-Rand permits a nonparametric estimation of treatment effect and hence is robust to the model specification as compared with difference-in-differences method\cite{Angrist2008, bertrand2004much}. Moreover, I-Rand enables us to draw inference on the significance of the estimated average treatment effect.
We demonstrate through simulations that the  I-Rand algorithm reduces error in estimates of the treatment effect compared to the pooled approach and difference-in-differences.

 The article is organized as follows. Section \ref{sec:methodology} introduces basic concepts from the potential outcomes model and matching methods, and propose the new I-Rand algorithm that we use to analyze the two-point time-series data. 
Section \ref{sec:compareofI-randwithothers} compares the proposed I-Rand with benchmark methods such as the pooled approach and the difference-in-differences. 
Section \ref{sec:diet} explains the use of I-Rand to understand the 
role of the LCD in reducing the risks of T2D and CVD risk. 
Section \ref{sec:obesity} investigates the relationship between obesity, T2D, and CVD risk. We discuss the limitations of our methods and indicate directions for future research in Section~\ref{sec:discussion}. 

\section{Motivation, Dataset, and Methodology}
\label{sec:methodology}

\subsection{A motivating example}
\label{sec:motivating}
\begin{figure}[!ht]
    \centering
    \includegraphics[scale=0.2]{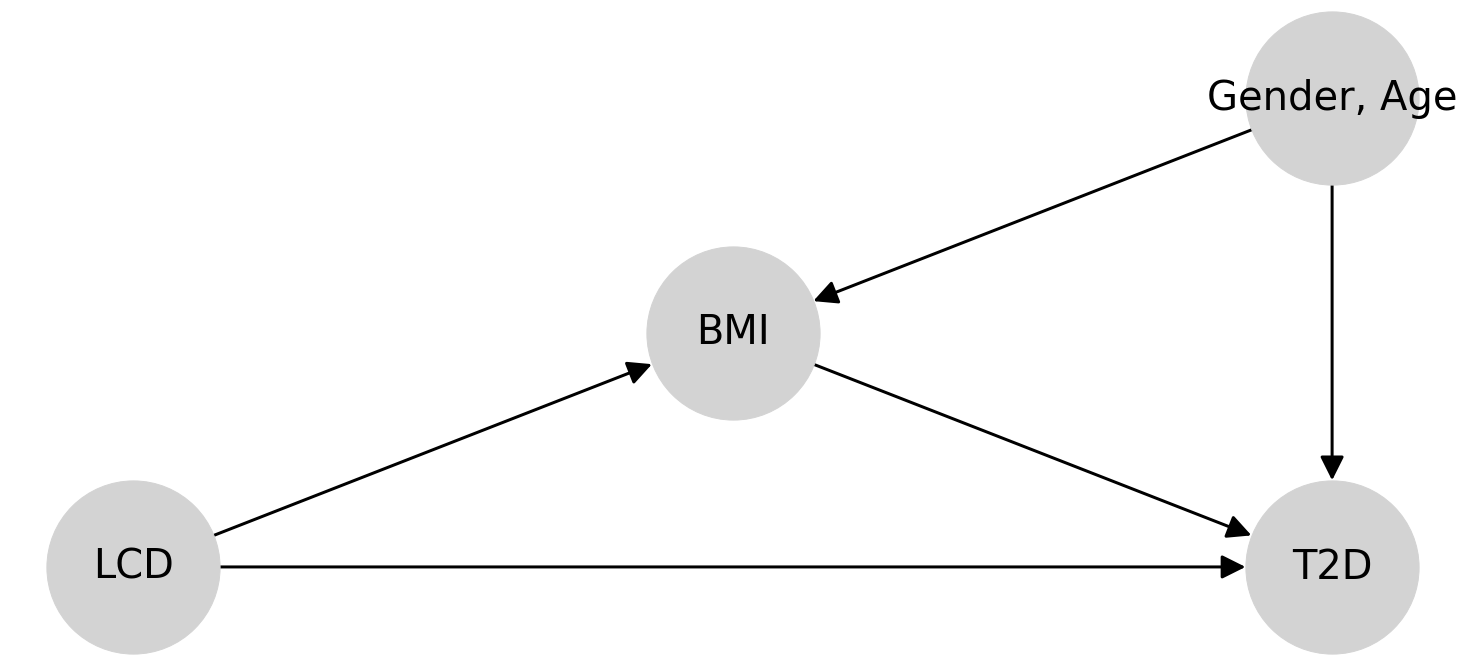}
    \caption{Assumed coarse-grained causal graph for the relationship between LCD, BMI, and the outcome T2D.}
    \label{fig:dag_lcd_t2d}
\end{figure}

Cause-and-effect questions arise naturally in the context of nutrition or health, making causal analysis especially relevant. 
Consider the counterfactual question, \emph{If an individual changes from a regular diet to an LCD, would he / she be less likely to develop T2D?} 
We can attempt to estimate the effect of diet on T2D from
observational data.  Any cause-and-effect inferences from observational data rely on restrictive assumptions and a specification of the underlying causal structure. In particular, we make the following assumptions.
First, the treatment is a binary variable that indicates whether or not an individual follows an LCD. 
\color{black}{The binary treatment LCD abstracts away the degree of LCD as this data there is clinical consultation }
Second, body mass index (BMI) is a surrogate for obesity and mediates the effect of LCD on T2D \cite{lavie2009}. Gender is a binary variable and  age is an ordinal variable.
Finally,  the medical outcome T2D is an ordinal variable indicating status at time of reporting: non-diabetics,  pre-diabetics, and  diabetics. T2D categories rely on glycated haemoglobin (HbA1c) value.
We also note that the BMI, age and gender variables  reflect only the \emph{case} demographics, i.e., the BMI, age and gender distributions among the \emph{tested individuals}, and not the general demographics. 
We assume the coarse-grained causal graph in Figure \ref{fig:dag_lcd_t2d}, and motivate it by thinking of the following data-generating process:
(1) LCD affects both BMI and the risk of T2D based on  established knowledge of causal effects in nutrition studies \cite{Bazzano2014, Halton2008, dekoning2011};
(2) Gender and age affect BMI and the risk of T2D, but not the treatment LCD;
(3) Conditional on the status of LCD, BMI, gender and age, T2D status is sampled as the medical outcome;
(4) There are no hidden confounders (i.e., causal sufficiency). We discuss the role of unobserved variables in Section \ref{sec:discussion}. 
We use arrows from one variable to another in the causal graph in Figure \ref{fig:dag_lcd_t2d} (and all other causal graphs) to indicate causal relationships.
Under these assumptions, we can estimate the effect of LCD on T2D  by adjusting for the confounders using the model of potential outcomes.

We will analyze the effect of LCD on the likelihood of developing T2D using Figure~\ref{fig:dag_lcd_t2d} after describing the structure of our dataset and reviewing causal inference basics.

\subsection{Data}
\label{subsec:data}

Our work is based on  routine clinical data concerning 256 patients collected between 2013 and 2019 at the Norwood General Practice Surgery in the north of England \cite{unwin2020insights}.  
As background, Norwood serves a stable population of approximately 9,800 patients, and an eight-fold increase in T2D cases was recorded over the last three decades.

Each patient visited the Norwood General Practice Surgery twice. The average time between visits was 23 months with a standard deviation of 17 months. 
\textcolor{black}{Each patient is offered to start an LCD subsequent to the first visit.}
\footnote{Conventional one-to-one general practice consultations  were used for LCD advice, supplemented by group consultation, to help patients better understand the scientific principles and consequences of LCD; including how glucose and insulin levels change in response to different foods \cite{unwin2020insights}. 
The role of group sessions was to reinforce diet and lifestyle change. LCD intervention encourages a reduction in the intake of sugary and starchy foods, for example, sugary breakfast cereals and rice, by replacing them with, for example, green leafy vegetables, eggs, meat  and fish, with sensitivity of each individual's socio-cultural dietary restrictions and preferences. }  Measurements of standard indicators such as age, gender, weight, HbA1c, lipid profiles, and blood pressure were taken at both visits. Since CVD includes a range of clinical conditions such as stroke, coronary heart disease, heart failure, and atrial fibrillation \cite{anderson1991}, several different risk factors are recorded for CVD during individuals' visits. 
We study four risk factors that indicate CVD risk.  These are systolic blood pressure, serum cholesterol level, high-density lipoprotein, and a widely used measure of CVD risk called the Reynolds risk score, which is designed to predict the risk of a future heart attack, stroke, or other major heart disease.
The Reynolds risk score  is a linear combination of different risk factors such as age, blood pressure, cholesterol levels and smoking habits \cite{ridker2007}\footnote{Some of the variables used in calculating the Reynolds risk score are missing from data. We make the simple choice of excluding them from the formula.}. A complete list of variables along with 
definitions and  summary statistics is in Appendix \ref{appendixA}.





\subsection{Model of potential outcomes}
\label{sec:potentialoutcome}
We use  concepts and notations from the Neyman (or Neyman-Rubin) model of potential outcomes \cite{neyman1990, rubin1974}.
The treatment assignment for individual $i$ is denoted by $T_i$, where $T_i=0$ and $T_i=1$ represent control and treatment. Let $Y_i$ be the observed outcome and $X_i$ be the observed confounders. For example, $X_i$ represents gender and age in the motivating example. 
The causal effect for individual $i$ is defined as the difference between the outcome if $i$ receives the treatment, $Y_i(1)$, and the outcome if $i$ receives the control, $Y_i(0)$. 
Since, in practice, an individual cannot be both treated and untreated, we work with 
two populations: a group of individuals exposed to the treatment and  a group of individuals  exposed to the control.
It is important to distinguish between the \emph{observed} outcome $Y_i$ and the \emph{counterfactual} outcomes $Y_i(1)$ and $Y_i(0)$. The latter are hypothetical and may never be observed simultaneously; however, they allow a precise characterization of questions of interest. 
For example, the causal effect for individual $i$ can be written as the difference in potential outcomes:
\begin{equation*}
\tau(X_i)  = \mathbb E[Y_i(1)|X_i]  -  \mathbb E[Y_i(0)|X_i].
\end{equation*} 
Since the outcome surface $\tau(X)$ depends on confounders, we focus on the ``average treatment effect" (ATE), $\mathbb E_X[\tau(X)]$, which is defined as the average causal effect for all individuals including both treatment and control. 

\begin{algorithm}
\caption{ \normalsize{Review of the propensity score matching algorithm}}\label{alg:psm}
\begin{algorithmic}[1]
\State  Define a distance measure for determining whether or not  an individual is a good match for another. For example, let the distance measure  $D_{ij}=|e_i - e_j|$, which is based on \emph{propensity score} $e_i(X_i) = P(T_i=1|X_i)$. We estimate $e_i$ by logistic regression for the case studies in Sections \ref{sec:obesity} and \ref{sec:diet}.
\State Given the distance measure, implement a matching method. For example, we apply matching with replacement and select a set of comparison units using the nearest-neighbor method in our case studies.
 Then we calculate ATE by
 \begin{equation*}
     \frac{1}{n}\sum_{i}\left(Y_i-\frac{1}{|J_i|}\sum_{j\in J_i}Y_j\right),
 \end{equation*}
 where $n$ is the sample size, $J_i$ is the set of individuals that belong to a different group (i.e., treatment or control group) than the individual $i$ and are matched to $i$, and $|\cdot|$ denotes the number of elements in the set.
\State Assess the quality of the matched samples and iterate with steps 1 and 2 until samples are well matched. Output ATE.
\end{algorithmic}
\end{algorithm}

Matching methods attempt to eliminate bias in estimating the treatment effect from observational data by balancing observed confounders across treatment and control groups; see, e.g., Rubin and Thomas \cite{rubin1996} and Imbens \cite{imbens2004}.
These works identify two assumptions on data that are required in order to apply matching methods in an observational study. 
\begin{itemize}
    \item \textcolor{black}{The \emph{strong ignorability condition} (Rosenbaum and Rubin \cite{rosenbaum1983central}) is referred to as the combination of exchangeability and positivity, which we discuss later that they are satisfied in our experiments.}
    \begin{itemize}
        \item Treatment assignment is independent of the potential outcomes given the confounders.
        \item There is a non-zero probability of receiving treatment for all values of $X$:
$0<\mathbb P(T=1|X)<1$.
    \end{itemize}
Weaker versions of the ignorability assumption exist; see, e.g., Imbens \cite{imbens2004}.
\item  The \emph{stable unit treatment value assumption} (SUTVA; Rubin \cite{rubin1980}), which states that the outcomes of one
individual are not affected by treatment assignment of
any other individuals.  There are two parts of the SUTVA assumption, which we rely on later in this paper.
    \begin{itemize}
         \item No-interference: The outcome for individual $i$ 
cannot depend on which treatment is given to individual $i'\neq i$. (Rubin \cite{rubin1980} attributes this to Cox \cite{cox1958}.)
         \item No-multiple-versions-of-treatment: There can be only one version of any treatment, as multiple versions might give rise to different outcomes. (Rubin \cite{rubin1980}  attributes this to Neyman \cite{neyman1935statistical}.)
    \end{itemize}
\end{itemize}
``Version"  refers to detailed information that is ignored as we coarsen a refined indicator to be used as a (typically binary) treatment.
The assumptions mentioned above are complementary to the assumptions that determine causal models such as the one shown in Figure~\ref{fig:dag_lcd_t2d}.  
To determine if treatment $T$ is ignorable relative to outcome $Y$, conditional on a set of matching variables, we  require only that matching variables block all the back-door paths between $T$ and $Y$, and that no matching variable is a descendent of $T$ \cite{pearl2009causality}. For example, LCD in Figure~\ref{fig:bmit2d} is ignorable since matching the confounders (i.e., gender and age) blocks all the back-door paths and the confounders are not descendants of LCD. 
The algorithm for propensity score matching is summarized in Algorithm \ref{alg:psm}.
Detailed discussions of each step are deferred to Appendix \ref{sec:matching}.

\subsection{I-Rand algorithm} \label{subsection:estimating_ate}

Two-point time-series datasets that are structurally similar to the nutrition dataset introduced in Section \ref{subsec:data} arise frequently in medical and health studies.  A dataset of this type consists of a baseline observation at time $t=0$ and a follow-up observation at $t=1$, where all individuals receive a treatment between the two time points.
How do we apply matching methods to estimate the causal effect of a treatment that was taken between the two time points from a dataset of this type?
To address this question, we look at what happens when we  attempt to apply statistical methods to estimate the causal effect. Although there are many popular machine learning methods for causal estimation \cite{chernozhukov2018double, dai2022orthogonalized}, we focus on two widely used approaches: pooling and difference-in-differences.


Pooling \cite{beck1995and, wilson2007lot} combines the baseline and the follow-up observations into a single  dataset. This approach treats the measurements from individual $i$ at $t=0$ (before taking the treatment) and $t=1$ (after observing the outcome of the treatment) as distinct data points. This amounts to using observations at $t=0$ as a control group.
Difference-in-differences \cite{Angrist2008, bertrand2004much}, on the other hand,
makes use of longitudinal data from both treatment and control groups to obtain an appropriate counterfactual to estimate causal effects. This approach compares the changes in outcomes over time between a population that takes a specific intervention or treatment (the treatment group) and a population that does not (the control group).

Consider the motivating example in Section~\ref{sec:motivating}, where every individual embarks on the  LCD treatment at time 0.
At time 1, we look at at how the outcome T2D is affected by the LCD between times 0 and 1, under numerous assumptions. Suppose we try to estimate the average treatment effect of the LCD  by  matching propensity scores on a dataset obtained by pooling observations at times 0 and 1.
Since, for every $i$, the treatment $T_{i,t}$ determines the treatment $T_{i, 1 - t}$ 
the outcome for individual $i$ at time $t$ depends on the treatment of individual $i$ at time $1-t$. In other words, the
pooled approach violates the no-interference assumption, and propensity score matching is not supported.\cite{beck1995and, wilson2007lot}.  As we illustrate with simulation in Section~\ref{sec:comtopoolunderlcd}, the no-interference violation can lead to sub-par performance of causal estimates based on pooling.
On the other hand, applying difference-in-differences to the motivating example would require us to make an assumption about what what would happen to individuals not treated between times $0$ and $1$.  We explore this in Section~\ref{sec:lcd-did}. 

The issues outlined above prompted us to develop I-Rand, a novel approach to estimating causal effects from two-point time-series data.
As we show in simulation, I-Rand can reduce estimation error introduced by violations of the SUTVA assumption incurred by pooling data.  
 There is some conceptional overlap between I-Rand and the \textit{synthetic control method} \cite{abadie2003synthetic, abadie2010synthetic}, which  provides a systematic way to choose comparison units (i.e., ``synthetic control'') as a weighted average of all potential comparison units that best resembles the characteristics of the unit of interest (i.e., treatment unit). In I-Rand, both the control and treatments units are chosen from the data to form a ``synthetic subsample'' from which the causal effect is estimated using propensity score matching (i.e., the one control unit with the closest propensity score to the treatment unit of interest).

I-Rand samples one of the two visits for each patient, calculates the ATE on this selected subsample, and shuffles the treatment of the subsample to estimate the significance of the treatment. The estimation relies on the matching method described in Section \ref{sec:potentialoutcome} and applies a permutation test to the statistics estimated from the matching methods on the subsamples to infer the significance. Under the null hypothesis, the empirical ATEs are identically distributed. Formally, we construct a subsample in which each patient appears exactly once, either at $t=0$ or $t=1$ with the same probability, and then  calculate the ATE from this sample. Then we construct additional $(M-1)$ subsamples, where each additional subsample should be drawn to have as few common observations with existing subsamples as possible.
For example, one can apply the Latin hypercube sampling \cite{mckay2000comparison} to draw the subsamples. 
We calculate the ATEs from the constructed $(M-1)$ subsamples and take the average ATE:
\begin{equation}
\label{eqn:I-Rand}
    \frac{1}{M}\sum_{m=1}^M \text{ATE}^{(m)},
\end{equation}
where $m$ indicates the $m$th generated subsamples. 
Then the i-Randomization estimator in Equation (\ref{eqn:I-Rand}) gives the overall estimated ATE.
To assess the significance of the treatment, we add another layer of randomization by permuting the treatments in the subsample. That is, given a subsample $m$ with corresponding estimand $\text{ATE}^{(m)}$, we shuffle the treatment vector of this subsample without changing the confounders or the outcome. We then estimate an average treatment effect $\text{ATE}^{(m, s)}$ for this shuffled treatment, where the superscript $(m, s)$ indicates that we have selected the subsample $m$  and the shuffle $s$. We repeat the experiment $S$ times (for a fixed subsample $m$), and obtain the distribution of average treatment effects., i.e., $(\text{ATE}^{(m, s)})_{s \in \{1, ..., S\}}$. Then we calculate a p-value as the fraction of permuted average treatment effects that exceed the estimand $\text{ATE}^{(m)}$. The additional complexity of I-Rand is justified by the benefits that it brings relative to the pooled approach and difference-in-differences. I-Rand overcomes the SUTVA violation that is inherent in the pooled approach, and it creates a synthetic control group, which is absent in difference-in-differences.
The I-Rand algorithm is summarized in Algorithm \ref{alg:i-rand}.

\begin{algorithm}
\caption{ \normalsize{I-Rand algorithm}}\label{alg:i-rand}
\begin{algorithmic}[1]
\State  \normalsize{\textbf{Input:} $2n\times p$  data matrix where each row is attributes of an individual $i\in\{1,\ldots,n\}$ at time point $\in\{0,1\}$, and $p$ is the number of variables including the treatment, confounder, and outcome.}
\State \textbf{for} $m=1,2,\ldots, M$ \textbf{do}
\State \quad\quad Sample a binary vector of length $n$, where the index is the individual's ID and the value is the time point (sampling without replacement). Select the corresponding subsample $m$;
\State \quad\quad Calculate $\text{ATE}^{(m)}$ by the matching method;
\State \quad\quad \textbf{for} $s=1,2,\ldots, S$ \textbf{do}
\State \quad\quad \quad\quad Shuffle the vector of treatment;
\State \quad\quad \quad\quad  Calculate $\text{ATE}^{(m, s)}$ for the shuffle $s$ of the treatment from subsample $m$;
\State \quad\quad \textbf{end for}
\State \quad\quad Calculate $\text{p-val}^{(m)} = \frac{1}{S}\sum_{s=1}^{S}\mathbbm{1}_{\text{ATE}^{(m, s)} > (\text{resp. } <) \text{ATE}^{(m)}}$. That is,  the p-value for the one-tailed test for the null hypothesis of no treatment effect.
\State \textbf{end for}
\State \textbf{Output:} The mean of ATEs $= \frac{1}{M}\sum_{m=1}^{M}\text{ATE}^{(m)}$; The mean of the p-values: $\frac{1}{M}\sum_{m=1}^{M}\text{p-val}^{(m)}$.
\end{algorithmic}
\end{algorithm}

We note that the permutation test in I-Rand is valid only if the rearranged data are exchangeable under the null hypothesis \cite{edgington2007}. 
In our two-sample test for the nutrition dataset, the exchangeability condition holds since the distributions of the two groups of data are the same under the null hypotheses that there is no treatment effect. 
The subsampling technique in I-Rand is similar to the one studied by Hahn \cite{Hahn2012} in the analysis of spatial point patterns.
The difference, however, is that the normalization of test statistics (i.e., ATE) is unnecessary in I-Rand since the matching method has  balanced the designs.

\section{Comparison of I-Rand with Alternative Methods}
\label{sec:compareofI-randwithothers}
We use simulation to compare errors in an I-Rand-based estimation of a treatment effect 
 with errors from
the pooled approach and difference-in-differences.
We look at causal effect estimation under two 
types of treatment assignments inspired by our data and the questions considered in this article. First,  we study the ``LCD-like treatment", as in the motivating example in Section \ref{sec:motivating}, where 
$T=0$ at $t=0$ and $T=1$ at $t=1$ for all individuals. The  LCD-like treatment respects the two-point time series structure since the assignment of $T$ depends on time.

Next, we consider a study from Section~\ref{sec:bmiandt2d}:  does obesity cause T2D?  Here, treatment is a binary indicator based on the body-mass index (BMI), where obesity is indicated by ${\rm BMI} > 30 $.  To avoid excess notation, we use the acronym ``BMI" to indicate both the body mass index and the binary treatment derived from it. In this study, there is a control group consisting of individuals with {\rm BMI} < 30. This treatment does not align with time, and we call treatments of this type ``BMI-like."\footnote{Practical considerations concerning the potential outcomes framework require that a treatment be a binary indicator, and that forces us to discard detailed information that may be contained by the continuous indicator BMI \cite{vanderweele2013causal}.} Here, it is natural to pool the data at the two time points, with a control group of non-obese individuals and a treatment group of obese individuals.  To apply difference-in-differences, we split the data into two subsets. The first subset consists of individuals who are non-obese at time 0. The control group in the subset is individuals who are non-obese at time 1, while the treatment group consists of individuals who are obese at time 1.   For this subset, the treatment, obesity, has a significant effect on T2D if change in T2D is significantly different in the treatment group than in the control group.   The second subset consists of individuals who are obese at time 1.  The control group in the subset is individuals who are obese at time 1, while the treatment group consists of individuals who are non-obese at time 1.  Again, the treatment, obesity, causes T2D if the change in T2D is significantly different from zero in the treatment group than in the control group. As usual, the numerous assumptions on which our results rely include causal completeness. We note that, while it may be unintuitive, it is certainly possible that the effect of increased obesity on T2D could turn out to be negative. An overview of the comparison of I-Rand with two benchmark methods is given in Table \ref{table:summaryofcomparison}.

\begin{table}
\caption{Overview of comparison of I-Rand with alternative methods given two-point time-series with novel structures}
\centering
\begin{tabular}{ l c c}
\toprule
 & Respect time structure & Ignore time structure\\
 & (LCD-like treatment) & (BMI-like treatment)\\
\midrule
Pooled approach vs. I-Rand & Section \ref{sec:comtopoolunderlcd} & Section \ref{sec:bmi-pooled} \\
Difference-in-differences vs. I-Rand & Section  \ref{sec:lcd-did} & Section \ref{sec:bmi-did}\\
\bottomrule
\end{tabular}
\label{table:summaryofcomparison}
\end{table}

All our simulations consider a panel dataset with two time points where outcomes are specified by the structural equation:
\begin{equation}\label{eq:generalmodel}
    \begin{aligned}
        Y_{i, t} & =  \alpha+f(T_{i, t}) + g(X_{i, t})+ \varepsilon^{Y}_{i, t},
    \end{aligned}
\end{equation}
where the confounder vector $X_{i,t}$, such as age or gender, takes continuous or categorical values. The parameter $\alpha\in\mathbb R$, and $f(\cdot)$ and $g(\cdot)$ are unknown functions that satisfy the condition $\mathbb E_T[f(T)]=\mathbb E_X[g(X)]=0$ to ensure identifiability of model (\ref{eq:generalmodel}). 
Assuming the confounder satisfies the back-door criterion \cite{pearl2009causality}, we can interpret $f(\cdot)$ as the causal mechanism of $T$ affecting $Y$ \cite{zhao2019causal}. The noise term $\varepsilon^{Y}_{i, t}$ is assumed to be i.i.d. for any $i$ and $t$, and has zero mean and bounded variance.  The treatment $T_{i, t}$  is specified differently in different examples that we consider below.

 
\subsection{Time-aligned (LCD-like) treatment} \label{sec:lcd-like}
To complete the specification of the data generating process (\ref{eq:generalmodel}), we set the treatment variable as follows:
\begin{equation}\label{eqn:LCD-treatment}
    T_{i,t} = \mathbbm{1}_{t = 1},\quad\forall i\in\{1,\ldots,n\} \text{ and }t\in\{0,1\},
\end{equation}
where the treatment $T_{i,t}$ for individual $i$ at time $t$ is binary and depends only on time. 
For example, $T_{i,t}$ in the nutrition data of Section \ref{subsec:data} indicates whether individual $i$ follows an LCD at time $t$.  The outcome $Y_{i,t}$ is analogous to the HbA1c measure in  the nutrition data of Section \ref{subsec:data}. \color{black}{
We note that the strong ignorability condition in Section \ref{sec:potentialoutcome} is satisfied under the LCD-like treatment \eqref{eqn:LCD-treatment}. Specifically, the first condition on exchangeability holds since the $T=1$ is independent of the potential outcomes given the confounders under \eqref{eqn:LCD-treatment}. The second condition on positivity holds because for any given confounders $X$ that excludes the time $t$, the probability of receiving treatment satisfies $0<\mathbb P(T=1|X)<1$.} \color{black}{However,  the data-generating process under \eqref{eqn:LCD-treatment} violates SUTVA in Section \ref{sec:potentialoutcome}.}




\subsubsection{Comparison to the pooled approach}
\label{sec:comtopoolunderlcd}




The pooled approach breaches the ``no-interference'' assumption as $T_{i,t}$ determines $T_{i,t'}$, where $t\neq t'\in\{0,1\}$. 
Thus, each pair of distinct observations has the same probability of being matched, which violates the ``no-interference'' assumption of the SUTVA in Section \ref{sec:potentialoutcome}. We refer readers to Appendix \ref{sec:matching} for an overview of the propensity score matching.

\begin{figure}[!ht]
    \centering
    \includegraphics[height=5cm, width=5.5cm]{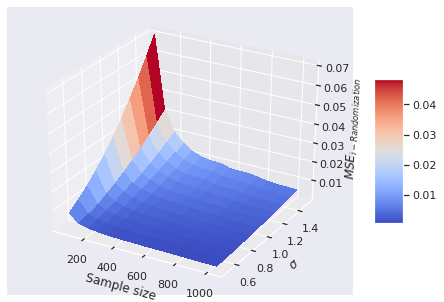}
    \includegraphics[height=5cm, width=5.5cm]{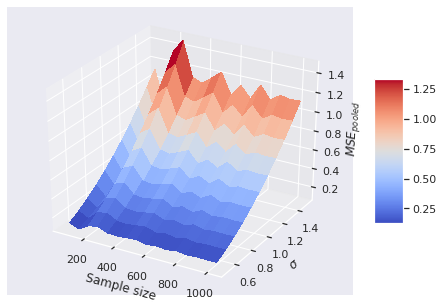}
    \includegraphics[height=5cm, width=5.5cm]{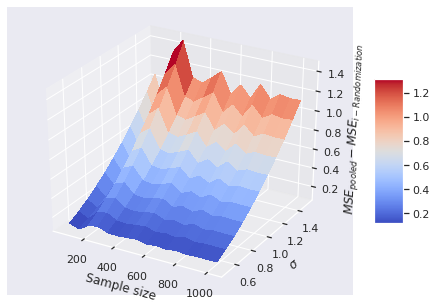}
    \caption{The  MSE for the estimate of treatment effect when varying the sample size and noise level $\sigma$.
    Left plot: the MSE surface for the I-Rand; Middle plot: the MSE surface for the pooled approach; 
    Right plot:  $\text{MSE(pooled)}-\text{MSE(I-Rand)}$.}
    \label{fig:simulation_pooled}
\end{figure}

We consider a numerical example that illustrates the consequence of breaching the ``no-interference'' assumption on the pooled data. We consider a correlated structure of confounders that simulates the age and gender in the nutrition data of Section \ref{subsec:data}, where 
\begin{equation}
\label{eqn:designofx}
\begin{aligned}
    X_{i, 0} & = \varepsilon_i^X,\quad \varepsilon_i^X \sim N(0,  1),\\
    X_{i, 1} & = \rho X_{i, 0} +  \sqrt{1 - \rho^2} \xi_{i}^X + a^Xt,\quad \xi_i^X \sim N(0,  1),
\end{aligned}
\end{equation}
where $\rho$ is the correlation between the confounder at $t=0$ and at $t=1$, and $a^X$ is a drift term. We set $\rho=1$ and $a^X=0$ in this simulation. It is the case when the confounder used is unchanged with the passage of time, i.e., fixed attributes like gender or ethnicity. The outcome $Y_{i,t}$ is generated by letting $f(\cdot)$ and $g(\cdot)$ in (\ref{eq:generalmodel}) be linear functions: \begin{equation}\label{eq:SEQ}
        f(T)  =  T\delta, \quad  g(X) = X\beta. 
\end{equation} 
Here $\alpha$ in (\ref{eq:generalmodel}) is set to $0$, and $\beta=-1$ and $\delta=1$ in (\ref{eq:SEQ}). The noise variable $\varepsilon^Y_{i, t}$ in (\ref{eq:generalmodel}) is independently drawn from $N(0, \sigma^2)$. 
Under the pooled approach, we estimate the treatment effect based on the propensity score matching in Algorithm \ref{alg:psm}. Under I-Rand, we estimate the treatment effect by averaging over the estimates using $100$ subsamples using Algorithm \ref{alg:i-rand}. Figure \ref{fig:simulation_pooled} reports the mean squared errors (MSEs) for $\delta$  with varied sample sizes and noise levels. In our example, I-Rand outperforms the pooled approach, 
whose ATE estimate has inflated error due to the breach of ``no-interference'' assumption.

\subsubsection{Comparison to difference-in-differences}
\label{sec:lcd-did}
The standard set up of difference-in-differences \cite{Angrist2008, bertrand2004much} is one where outcomes are observed
for two groups for two time periods. One of the groups is exposed to a treatment in the second
period but not in the first period. The second group is not exposed to the treatment during
either period. In the case where the same units within a group are observed in each time period,
the average gain in the control group is substracted from the average gain in the treatment group, which gives an estimate to the average treatment effect:
\begin{equation}
\label{eqn:nonparametricdid}
\begin{aligned}
   \text{ATE}&\equiv \mathbb E_X[\mathbb E[Y_i(t=1)-Y_i(t=0)|T_i(t=1)=1,T_i(t=0)=0,X]]\\
   &\quad - \mathbb E_X[\mathbb E[Y_i(t=1)-Y_i(t=0)|T_i(t=1)=0,T_i(t=0)=0,X]].
\end{aligned}
\end{equation}
Difference-in-differences removes biases in second period comparisons between the treatment
and control group that could be the result from permanent differences between those groups, as
well as biases from comparisons over time in the treatment group that could be the result of trends. 
We note that this standard difference-in-differences approach does not require the knowledge of the functions $f(\cdot)$ or $g(\cdot)$ in (\ref{eq:generalmodel}).
However, in our application with the LCD-like treatment design (\ref{eqn:LCD-treatment}), the  treatment effect (\ref{eqn:nonparametricdid}) cannot be estimated from data using the aforementioned standard approach of difference-in-differences.
The main reason is that LCD-like treatment design lacks 
the control group $\{i|T_i(t=1)=0,T_i(t=0)=0\}$.
We summarize this result in the following theorem.
\begin{theorem}
\label{prop:nonparaofdid}
Under the two-point treatment design (\ref{eqn:LCD-treatment}) and the structural equation (\ref{eq:generalmodel}), the treatment effect (\ref{eqn:nonparametricdid}) is not identifiable by difference-in-differences if there is no prior knowledge on the parametric family of  $f(\cdot)$ and $g(\cdot)$ in (\ref{eq:generalmodel}).
\end{theorem}
\noindent
The proof of Theorem~\ref{prop:nonparaofdid} is in Appendix~\ref{app:proofs}. The theorem suggests that the difference-in-differences is not applicable under the two-point treatment design (\ref{eqn:LCD-treatment}). By contrast, the I-Rand in Algorithm \ref{alg:i-rand} constructs a synthetic control group by randomly selecting one of the two points for each individual, and it can identify and estimate the treatment effect (\ref{eqn:nonparametricdid}) from data.

A remedy for applying difference-in-differences to the treatment design (\ref{eqn:LCD-treatment}) is assuming a parametric functional form of $f(\cdot)$ and $g(\cdot)$ in the structural equation (\ref{eq:generalmodel}). For example,  one can assume $f(\cdot)$ and $g(\cdot)$ to be linear functions (\ref{eq:SEQ}). 
The treatment effect $\delta$ can be estimated by regressing over the observational treatment group data $\{i|T_i(t=1)=1,T_i(t=0)=0\}$ under the design (\ref{eqn:LCD-treatment}).
Nonetheless, we demonstrate that even in a parametric structural equation,
I-Rand can outperform difference-in-differences. 
We simulate data using formulas  (\ref{eq:generalmodel}) to  (\ref{eq:SEQ}).
The outcome $Y_{ij}$ is calculated using (\ref{eq:SEQ}) with $\alpha=0$, $\beta=-1$, and $\delta=1$, whereas the errors $\varepsilon^Y_{i, t}$ are i.i.d  drawn from $N(0, \sigma^2)$. 
In (\ref{eqn:designofx}), 
we set $\rho=0.99$ and $a^X=1/12$, leading to similar values of the covariates for each individual at the two time points as the nutrition data in Section \ref{subsec:data}, and a passage of time equal to one month. We take the difference in the variables in  both sides of the equation (\ref{eq:generalmodel}) and obtains
    \begin{equation}\label{eqn:difference}
        DY_i = DT_{i}\delta + DX_{i}\beta +  D\varepsilon_{i}^Y,
    \end{equation}
where the operator $D$ denotes the difference in the variable between $t=1$ and $t=0$, i.e., $DZ_i = Z_{i, t=1} -  Z_{i, t=0}$ for any variable $Z$. 
In this example,
the propensity score matching is not applicable to difference-in-differences \footnote{The difference-in-differences fails to meet the strong ignorable treatment assignment condition in Section \ref{sec:potentialoutcome}. Specifically, $ 0<P(\text{Treatment}=1|X)<1$, as $P(DT=1|X)=1$  and $P(DT=0|X)=0$.
Hence we cannot directly apply the propensity score matching in Section \ref{sec:matching} to estimate the treatment effect.}.
We apply least squared  regression to model (\ref{eqn:difference}) and use the observational treatment group data $\{i|T_i(t=1)=1,T_i(t=0)=0\}$ to estimate the parameter $\delta$. 
For I-Rand, we apply Algorithm \ref{alg:i-rand} and obtain the treatment effect by averaging over $100$ subsamples. Difference-in-difference estimates are obtained by least squared regression on (\ref{eqn:difference}), as explained above.

\begin{figure}[!ht]
    \centering
    \includegraphics[height=5cm, width=5.5cm]{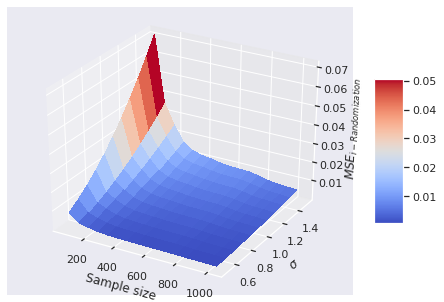}
    \includegraphics[height=5cm, width=5.5cm]{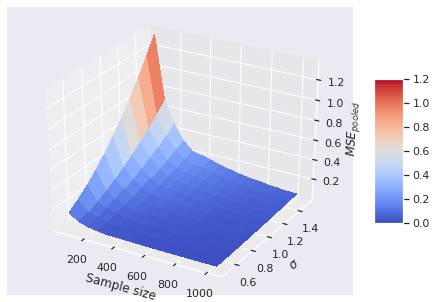}
    \includegraphics[height=5cm, width=5.5cm]{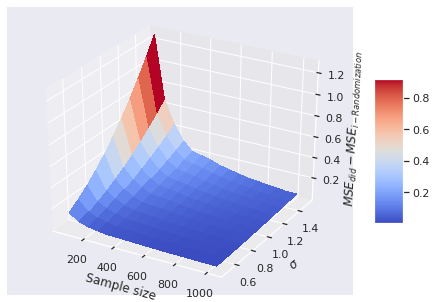}
    \caption{  The  MSE for the estimate of treatment effect when varying the sample size and noise level $\sigma$.
    Left plot: the MSE surface for the I-Rand; Middle plot: the MSE surface for the DiD approach; 
    Right plot:  $\text{MSE(difference-in-differences)}-\text{MSE(I-Rand)}$.}
    \label{fig:simulation_did}
\end{figure}

Figure \ref{fig:simulation_did} reports the mean-squared errors to the true value $\delta=1$ (left panel), and the difference in MSEs between i-Rand and DiD, when varying sample sizes and noise levels.  From the plots, we see that i-Rand has small MSEs and that
MSEs for DiD are larger than MSEs for i-Rand. While  linear regression is subject to large estimation error because of the high correlation between change in the treatment vector ($DT$) and  change in age ($D X$), the absence of a control group ($DT = 0$) makes the passage of time (increase in age) as likely to be responsible for the change in outcome as the treatment itself. While the poor performance of difference-in-differences can be traced to the lack of a control group, an estimation using regression does not always fail and could still give good results when the treatment and confounders are uncorrelated. Our example illustrates how regressions can fail to estimate a causal effect. 




We summarize in Table \ref{table:comparison}   the advantages of  I-Rand compared to two benchmark approaches for the LCD-like treatment. 
\begin{table}
\caption{Comparison of three approaches in the case of the LCD-like treatment in Section \ref{sec:lcd-like}}
\centering
\begin{tabular}{ l c c c}
\toprule
 & Pooled approach & Difference-in-differences & I-Rand\\
\midrule
SUTVA assumption & Fail & Hold & Hold\\
Control group & Yes & No & Yes\\
\bottomrule
\end{tabular}
\label{table:comparison}
\end{table}

\subsection{Time misaligned (BMI-like) treatment}

To complete the specification of the data generating process (\ref{eq:generalmodel}), we set the treatment variable as follows:
\begin{equation}\label{eqn:bmitreatment}
    T_{i,t} = h(X_{i,t}),\quad \forall i\in\{1,\ldots,n\}\text{ and } t\in\{0,1\}.
\end{equation}
Here the treatment $T_{i,t}$ for individual $i$ at time $t$ is a binary function of the confounders $X_{i,t}$.

Our treatment is time misaligned because it ignores our two-point time-series structure, i.e., two observations for each patient with treatment administrated at $t=0$ and observed at $t=1$.  It mimics the experiment  in Section \ref{subsec:data}, where the treatment is a discrete version of BMI:  $T_{i,t}$ is weight category (e.g., normal or overweight) of individual $i$ at time $t$. In this experiment $T_{i,t}$ depends on the confounders such as LCD, age and gender.

\subsubsection{Comparison to the pooled approach}
\label{sec:bmi-pooled}

Unlike the LCD-like assignment in Section \ref{sec:comtopoolunderlcd}, the pooled approach \cite{beck1995and, wilson2007lot} meets the SUTVA assumption of ``no-interference'' under  design (\ref{eqn:bmitreatment}). Moreover, the pooled approach provides an estimate to $\text{ATE}^*$ in (\ref{eqn:avegoal}) by treating the observations from an individual at $t=0,1$ as two distinct data points.  
We demonstrate through numerical examples that the pooled approach is a comparable alternative to I-Rand in the BMI-like treatment assignment (\ref{eqn:bmitreatment}). 

\begin{figure}[!ht]
    \centering
    \includegraphics[height=5cm, width=5.5cm]{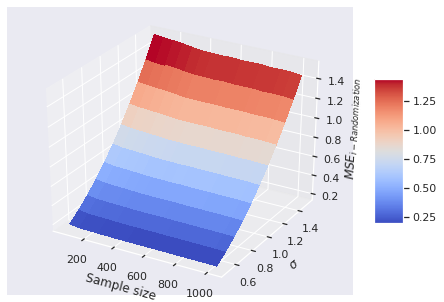}
    \includegraphics[height=5cm, width=5.5cm]{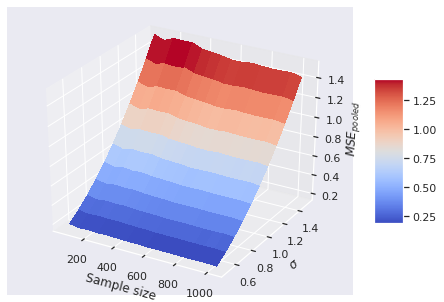}
    \includegraphics[height=5cm, width=5.5cm]{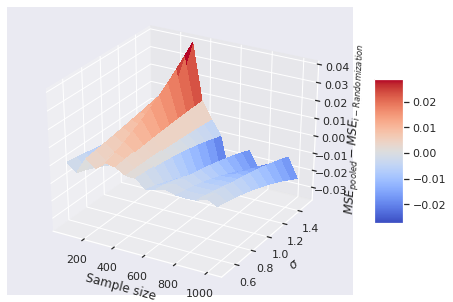}
    \caption{The MSE for the estimate of treatment effect when varying the sample size and noise level $\sigma$.
    Left plot: the MSE surface for the I-Rand; Middle plot: the MSE surface for the pooled approach; 
    Right plot:  $\text{MSE(pooled)}-\text{MSE(I-Rand)}$.}
    \label{fig:simulation_bmi_pooled}
\end{figure}

We specify the confounder $X=(X^{(1)},X^{(2)})$, where $X^{(1)}$ denotes an individual's age and $X^{(2)}$ indicates whether or not an individual has followed an LCD. Age, $X^{(1)}$, follows the confounder generating process
in (\ref{eqn:designofx}) with correlation $\rho=0.99$  and passage of time $a^X=1/12 = {\rm one~month}$.  (The parameter $a^X$ in (\ref{eqn:designofx}) is analogous to the time trend in the nutrition data in section \ref{subsec:data}.) 
The LCD-indicating confounder $X^{(2)}$ is set to $X^{(2)}_{i, t}=\mathbbm{1}_{t = 1}$. Treatment is assigned according to $T_{i,t} = \mathbbm{1}_{\varepsilon^T_{i, t}>0}$ where $\varepsilon^T_{i, t} = X^{(2)}_{i, t} + \xi^T_{i, t}$ where $\xi_{i,t}^T \sim N(0, 1)$. 
We consider the linear model (\ref{eq:SEQ}) for outcome $Y_{i,t}$, where $\varepsilon_{i, t}^{Y} \sim N(0, \sigma^2)$, and  $\alpha=0$, $\beta=(-1, 1)$, and $\delta=1$. 
The results are displayed in Figure \ref{fig:simulation_bmi_pooled}, where the MSE surface for I-Rand is shown with varying sample size and noise level $\sigma$.  Also displayed is the difference between the MSE of I-Rand and the pooled approach.  I-Rand performs similarly to the pooled approach in our simulation, lending support to the assertion that the approaches are comparable for the BMI-like treatment (\ref{eqn:bmitreatment}). 

\subsubsection{Comparison to difference-in-differences}
\label{sec:bmi-did}
In the time misaligned BMI-like treatment, difference-in-differences \cite{Angrist2008, bertrand2004much} encounters the problem of having four different types of individuals; always-treated ($\{i|T_i(t=1)=1,T_i(t=0)=1\}$), never-treated ($\{i|T_i(t=1)=0,T_i(t=0)=0\}$), treated-to-untreated  $\{i|T_i(t=1)=0,T_i(t=0)=1\}$, and untreated-to-treated $\{i|T_i(t=1)=1,T_i(t=0)=0\}$. To obtain an estimate of the treatment effect in this case, it is necessary to compare the outcomes of the group of never-treated to untreated-to-treated or the outcomes of the group of always-treated to treated-to-untreated. The idea is that the treatment state should be the same in both groups at $t=0$ and different at $t=1$.
We illustrate our ideas on the former;  the latter follows the same line of reasoning.
Difference-in-differences gives an estimate of the causal effect in (\ref{eqn:nonparametricdid}), which is the same as the target effect $\text{ATE}^*$ in (\ref{eqn:avegoal}) only if  
\begin{align}
   &\mathbb E_X[\mathbb E[Y_i(t=0)|T_i(t=1)=1,T_i(t=0)=0,X]]
    = \mathbb E_X[\mathbb E[Y_i(t=0)|T_i(t=1)=0,T_i(t=0)=0,X]],\label{eqn:cond1}\\
    &\text{ or }\quad \mathbb E_X[\mathbb E[Y_i(t=1)|T_i(t=1)=0,T_i(t=0)=0,X]]
    = \mathbb E_X[\mathbb E[Y_i(t=0)|T_i(t=1)=0,T_i(t=0)=0,X]].\label{eqn:cond2}
\end{align}
However, both (\ref{eqn:cond1}) and (\ref{eqn:cond2}) are strict and likely to fail in practice. 
Take the nutrition data in Section \ref{subsec:data} as an example.  Condition (\ref{eqn:cond1}) requires that the expected outcome at the baseline is  the same between two different groups: $\{i|T_i(t=1)=1,T_i(t=0)=0\}$ and $\{i|T_i(t=1)=0,T_i(t=0)=0\}$. However, the unobserved confounders such as lifestyle and genetic information in the two groups $\{i|T_i(t=1)=1,T_i(t=0)=0\}$ and $\{i|T_i(t=1)=0,T_i(t=0)=0\}$ are different (otherwise the treatment at $t=1$ should be the same in two groups), so that condition (\ref{eqn:cond1}) is likely to fail. Moreover, condition (\ref{eqn:cond2}) requires the expected outcomes be the same at the two time points, $t=0,1$, for the group $\{i|T_i(t=1)=0,T_i(t=0)=0\}$. However, since an individual does not take an LCD at $t=0$ and does take  an LCD at $t=1$, the confounder LCD assignment differs between $t=0$ and $t=1$. Hence the condition (\ref{eqn:cond2}) would fail for the nutrition data  in Section \ref{subsec:data}. 

\begin{figure}[!ht]
    \centering
    \includegraphics[height=5cm, width=5.5cm]{files/graphs/simulation/MSE_iRand_BMI.png}
    \includegraphics[height=5cm, width=5.5cm]{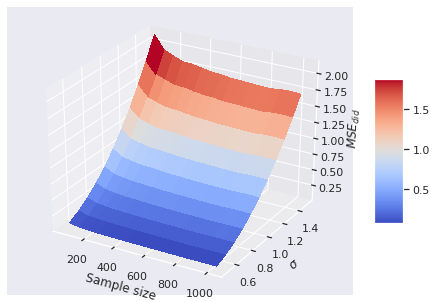}
    \includegraphics[height=5cm, width=5.5cm]{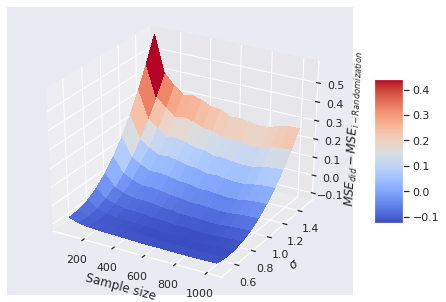}
    \caption{The MSE for the estimate of treatment effect when varying the sample size and noise level $\sigma$.
    Left plot: the MSE surface for the I-Rand; 
    Middle plot: the MSE surface for the DiD approach; 
    Right plot:  $\text{MSE(difference-in-differences)}-\text{MSE(I-Rand)}$.}
    \label{fig:simulation_bmi}
\end{figure}

Consequently,difference-in-differences (\ref{eqn:nonparametricdid}) cannot be applied to the BMI-like treatment assignment (\ref{eqn:bmitreatment}). 
An alternative approach to implementing difference-in-differences  is directly applying regression to a difference of parametric structural equation models. 
Specifically, we apply regression to (\ref{eqn:difference}). We consider the simulation setup in Section \ref{sec:bmi-pooled} and find that the estimate of the causal effect with difference-in-differences is unstable. 
The results are in Figure \ref{fig:simulation_bmi}, where the MSE surface for I-Rand is shown with varying sample size and noise level $\sigma$.  Also shown is the difference between the MSE of I-Rand and the MSE of the benchmark, difference-in-differences. We notice in the small sample size regime, I-Rand outperforms difference-in-differences. 

To conclude this section, we  stress that the first argument in favor of the application of I-Rand is its verification of the SUTVA assumption 
in both the time-aligned and time-misaligned treatments we considered. The estimation of the causal effect is 
data dependent, but we find that I-Rand performs at least as 
well as the benchmark methods in the examples considered.
Naturally, I-Rand is also subject to some limitations. One of them is the dependence of the estimates across
subsamples
inflate the variance estimate.
This can be mitigated
by minimizing overlaps in subsamples.
\section{Case Study I: Can Diet Lower the Risk for T2D and CVD?}
\label{sec:diet}

\subsection{Treatment effect of LCD on T2D}
\label{sec:lcdandt2d}
We can now analyze the motivating example introduced in Section \ref{sec:motivating} and give an answer to the counterfactual question: \emph{If an individual changes from a regular diet to an LCD diet, would he / she be less likely to develop T2D?} 
The LCD restricts  consumption of carbohydrates relative to the average diet \cite{Bazzano2014}.  
Several systematic reviews and meta-analyses of randomized control trials suggest beneficial effects of LCD in T2D and CVD, including improving glycaemic control, triglyceride and HDL cholesterol profiles \cite{meng2017efficacy, gjuladin2019effects, van2018effects}. However, the impact of LCD in a ``real world" primary care setting with observational data and its cause-and-effect inferences has not been fully evaluated \cite{unwin2020insights}. 
The challenges of analyzing routine clinical data include the irregular treatment assignments.
For example, our analysis relies on the two-point time-series data without control group described in Section \ref{subsec:data}, where all patients participated in the program are suggested to change from their regular diets to LCD after their initial visit to the clinic.
The irregular design of treatments limit the applications of benchmark methods such as pooled approach and difference-in-differences as discussed Section \ref{sec:compareofI-randwithothers}. In this section, we apply the proposed I-Rand algorithm to analyze the real data described in Section \ref{subsec:data}. 

The analysis using observation data utilizes the model of potential outcomes in Section \ref{sec:potentialoutcome}.
According to the causal graph in Figure \ref{fig:dag_lcd_t2d}, LCD takes the role of a treatment that affects the mediator BMI and outcome T2D. Gender and age affect BMI and T2D, but not the treatment LCD.  To quantify the expected change in T2D if BMI were changed, we need to calculate the total causal effect of LCD on T2D, which can be characterized by the ATE:
\begin{equation*}
    \mathbb E[\tau_1(\text{Gender}_i,\text{Age}_i)],
\end{equation*}
where the potential outcome $\tau_1$  (with ``$1$" indexing that this is the first of a series of nutrition questions)  is defined as
\begin{equation*}
    \tau_1(\text{Gender}_i,\text{Age}_i) = \mathbb E[\text{T2D}_i(\text{LCD}=1)\ |\ \text{Gender}_i,\text{Age}_i]  -  \mathbb E[\text{T2D}_i(\text{LCD}=0)\ |\ \text{Gender}_i,\text{Age}_i].
\end{equation*}
We control for the confounders (i.e., gender and age) \cite{pearl2009causality} to estimate the ATE and assess the significance by the proposed I-Rand algorithm.
We implement I-Rand by drawing $500$ subsamples and calculate the ATE of each subsample. Then, we perform the permutation test for each subsample to evaluate the significance level of the ATE.
The result provided in Table \ref{table:maoflcdoncvd} indicates that LCD would significantly decrease in the risk of T2D, which is also supported by the box plot of p-values in the first row of Figure \ref{fig:baratelcd}, and the distributions of ATEs and p-values in Appendix \ref{sec:lcdt2d}, where the results show the consistency of the significant causal effects across random subsamples. 
We make four remarks on the application of I-Rand and the experimental results of this example.

First, there is no control group with individuals on a regular diet at two visits. This is because all individuals were at risk of developing T2D or with T2D and thus suggested to begin the LCD after their first visit.
The application of the I-Rand algorithm in this example not only avoids a violation of the SUTVA assumption, but more importantly, 
to artificially construct synthetic control group. The way that I-Rand constructs synthetic control group is different from the existing synthetic control method \cite{abadie2010synthetic}. In particular, existing synthetic control method requires the available control individuals and constructs a synthetic control as a weighted average of these available control individuals. However, I-Rand does not require that there exists available control individuals. Instead,  I-Rand  constructs a synthetic control by subsampling one of the two time points of each individual.

\begin{table}
\caption{Causal analysis for the effect of LCD on T2D and Reynolds risk score for CVD}
\centering
\begin{tabular}{ c c c c c}
\toprule
 & LCD on T2D ($\mathbb E[\tau_1]$) &  & LCD on Reynolds risk score for CVD &\\
\cline{3-5}
 & & Total effect ($\mathbb E[\tau_2]$)  & Direct effect ($ \mathbb E[\tau_3]$) & Indirect effect ($ \mathbb E[\tau_4]$)\\
 \midrule
ATE & -0.593 & -0.015 & -0.009 & -0.005\\
p-value & 0.001 & 0.024 & 0.107 & 0.003\\
\bottomrule
\end{tabular}
\label{table:maoflcdoncvd}
\end{table}

Second, we note that under the null hypothesis of no causal effect,
the p-values follow a uniform distribution on $(0,1)$ given sufficiently many subsamples. However, the box plot of p-values in the first row of Figure \ref{fig:baratelcd}, corresponding to  the causal graph in Figure \ref{fig:dag_lcd_t2d}, shows p-values are concentrated at the origin, which indicates a strong evidence for the alternative hypothesis. We note that the hypothesis testing is performed for each subsample independently, but the p-values are not  independent across subsamples. This is because the subsamples are correlated although the correlation is weak  given each subsample is randomly chosen from the pool of $2^{256}$ subsamples. If the concentration of the p-values is around $0$, we can say with confidence that a small p-value is not a coincidence of the subsample, if most p-values are large, we conclude that the significance of the treatment effect is questionable.

Third, for better appreciating the results in Table \ref{table:maoflcdoncvd} we compare them with T2D risks from routine care without LCD suggestion. Some idea of the results that one might expect from routine care can be drawn from the data of control group in the DiRECT study \cite{lean2018primary}, which recently investigated a very low-calorie diet of  less than $800$ calories and subsequent drug-free improvement in T2D, including T2D remission without anti-diabetic medication. At 12 months, DiRECT study gives $46\%$ of T2D remission, which is close the $45\%$ rate given in Table \ref{table:maoflcdoncvd} from our dataset with LCD over an average of 23 months duration. As a comparison, DiRECT quotes a remission rate at 24 months of just $2\%$ for routine T2D care without dietary suggestion. This result emphasizes how rare remission is in usual care and the potential value of LCD to lower the T2D risk.

Finally, we note that our approach relies on individuals' assertions of compliance to the LCD.
For several years an LCD has generally been accepted as one containing less than 130 grams of carbohydrate per day \cite{accurso2008dietary}. However, it may not be realistic for individuals to count grams of carbohydrate in a regular basis. Our dataset collected from Norwood general practice surgery instead only give clear and simplified explanations of how sugar and carbohydrate affect glucose levels and how to recognize foods with high glycaemic loads \cite{unwin2020insights}. The promising result in Table \ref{table:maoflcdoncvd} shows that this simple and practical approach to lowering dietary carbohydrate leads to significant improvement in T2D without the need for precise daily carbohydrate or calorie counting.

\subsection{Mediation analysis for the effect of LCD on CVD}
\label{sec:medationlcdoncvd}

Motivated by the fact that T2D was crucial in explaining CVD risk (Benjamin et al. \cite{benjamin2018}), we seek to understand the role of T2D as a mediator of the effect of dietary on CVD risk. This is relevant from the perspective of clinical practice for an individual who is afflicted with both T2D and CVD, since he / she may be able to control factors besides T2D that contribute to CVD risk.  

\subsubsection{Causal graph of T2D as a mediator} 
\begin{figure}[!ht]
    \centering
    \includegraphics[scale=0.2]{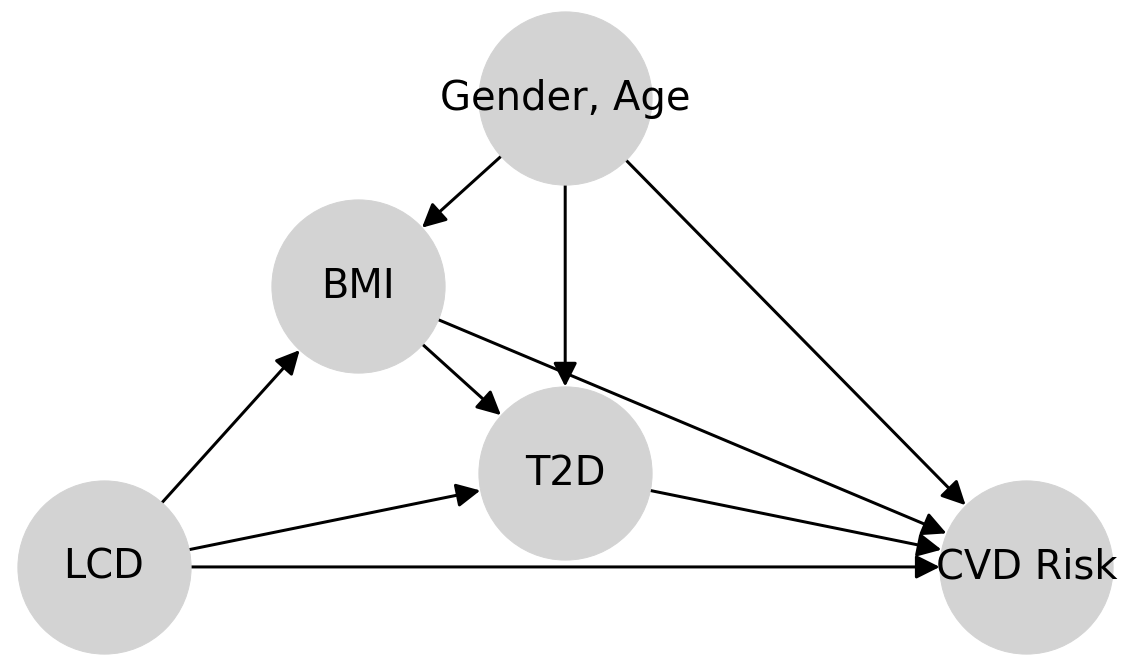}
    \caption{Assumed coarse-grained causal graph for the relationship between LCD, BMI, T2D, and the outcome CVD risk. Within this view, T2D acts as a mediator of the effect of LCD on CVD risk.}
    \label{fig:dag_lcd_cvd2}
\end{figure}

We assume the causal graph in  Figure  \ref{fig:dag_lcd_cvd2}. Note that the outcome CVD  has many risk factors, including systolic blood pressure, serum cholesterol level, high-density lipoprotein (which is inversely correlated with CVD risk); see, e.g., Ridker et al.  \cite{ridker2007}. We study these three well-known risk factors as well as the Reynolds risk score.
We motivate Figure  \ref{fig:dag_lcd_cvd2} with the following data-generating process: (1) Similar to Figure \ref{fig:dag_lcd_t2d}, choose the treatment LCD at random; Given a selected LCD, sample an individual with a corresponding BMI level; Conditional on the choice of LCD and BMI level, sample the T2D status as the medical outcome;
(2) In addition to Figure \ref{fig:dag_lcd_t2d}: Conditional on the choice of LCD and T2D status, sample the medical outcome within a given CVD risk factor. The details are as follows.
{\color{black}First, the arrows LCD $\to$ T2D and LCD $\to$ BMI encode that the distributions of T2D and BMI depend on LCD status. This dependence was quantified in Section \ref{sec:lcdandt2d}. 
Second, the arrow T2D $\to$ CVD reflects the established knowledge in nutrition science that T2D influences  CVD risk (Benjamin et al. \cite{benjamin2018}, Martín-Timón et al. \cite{ Mar2014}). Likewise, the arrow BMI $\to$ CVD translates the fact that obesity is a cardiovascular risk factor (Sowers \cite{sowers2003}). }
Finally,  since our model assumes causal sufficiency, the arrow LCD $\to$ CVD represents dietary-specific influences on CVD risk.  In reality, there may be other mediators, such as socioeconomic status, culture occupation, and stress level. 
In addition to the causal graph in Figure \ref{fig:dag_bmi_cvd2}, we assume there are no hidden confounders. 

Given these assumptions, we see that LCD causally influences CVD risk along two different paths:
 a path LCD $\to$ CVD, giving rise to a \emph{direct effect}, and two paths LCD $\to$ BMI $\to$ T2D $\to$ CVD and LCD $\to$ T2D $\to$ CVD, which are mediated by T2D and give rise to an \emph{indirect effect}.
Note that the direct effect of LCD on CVD risk is likely mediated by additional variables that are subsumed in LCD $\to$ CVD. We discuss this point further in Section \ref{sec:discussion}.
In mediation analysis, the goal is to quantify direct and indirect effects. 
We start with the total effect and then formulate the direct and indirect effects by allowing the treatment to propagate along one path while controlling the other path. 

\subsubsection{Total effect of LCD on CVD}

\begin{figure}[!ht]
    \centering
    \includegraphics[width=0.9\textwidth]{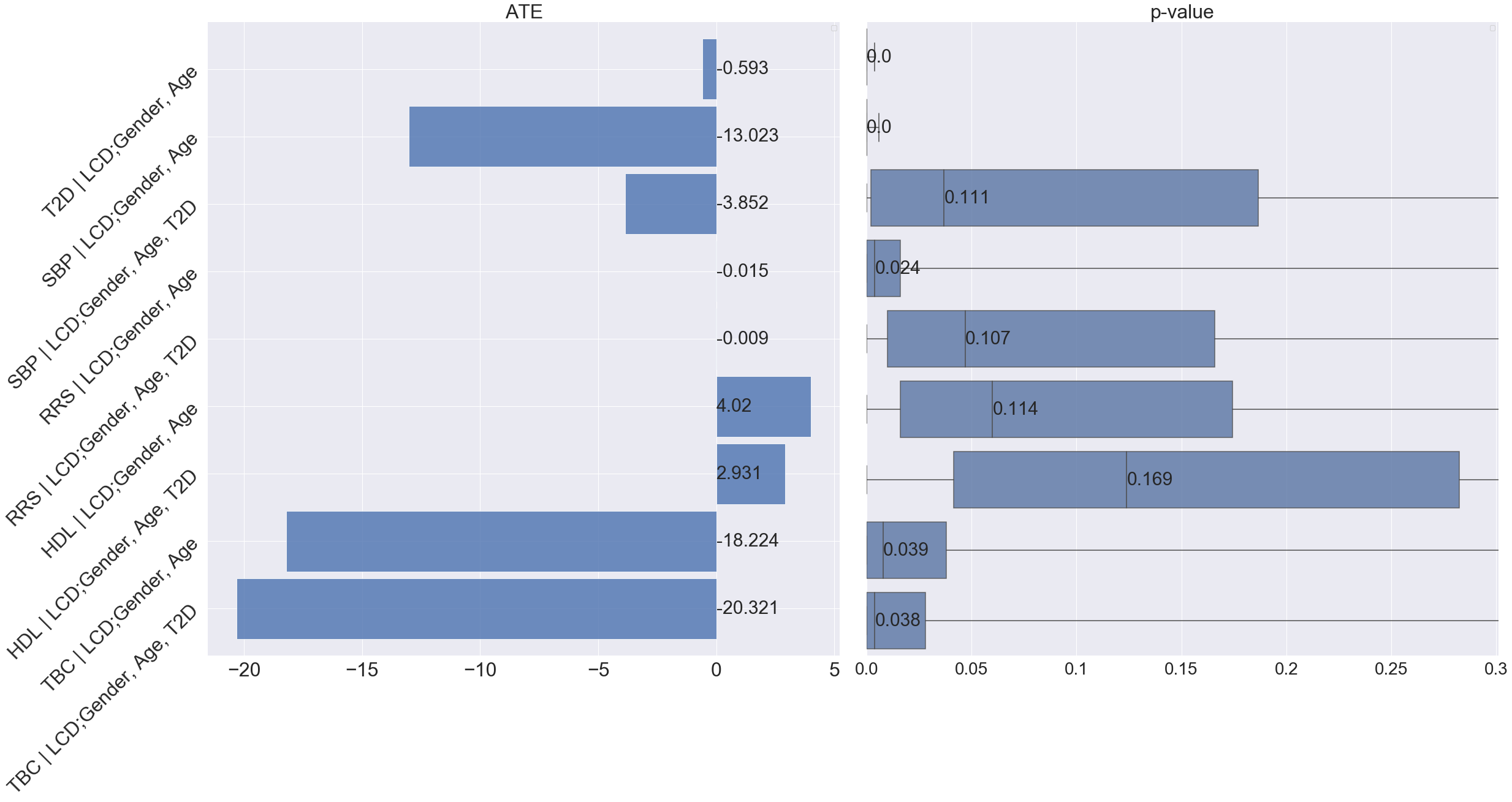}
    \caption{Mean ATE bar plot (left) and p-values box plot (right) for LCD as the treatment.  Each row corresponds to a causal diagram:“Outcome $|$ Treatment; Confounders”.  For example, “SBP $|$ LCD; Gender, Age” represents the causal diagram with the systolic blood pressure as the outcome, and gender and age as the confounders, and the LCD as the treatment.}
    \label{fig:baratelcd}
\end{figure}

Given the causal assumptions in the previous section, the first measure  of interest is the total causal effect of LCD on CVD, i.e., the answer to the following question:
\begin{quote} \emph{``What would be the effect on CVD if an individual changes from regular diet to LCD?"}
\end{quote}
We formulate the answer using the ATE:
\begin{equation*}
    \mathbb E[\tau_2(\text{Gender}_i,\text{Age}_i)],
\end{equation*}
where the potential outcome $\tau_2$ is defined as
\begin{equation*}
    \tau_2(\text{Gender}_i,\text{Age}_i) = \mathbb E[\text{CVD}_i(\text{LCD}=1)|\text{Gender}_i,\text{Age}_i]  -  \mathbb E[\text{CVD}_i(\text{LCD}=0)|\text{Gender}_i,\text{Age}_i].
\end{equation*}
Using the I-Rand algorithm, we report the results for the effect of LCD on the Reynolds risk score as measure of CVD risk. The total effect and the p-value are given in Table \ref{table:maoflcdoncvd}.
Figure \ref{fig:baratelcd}  summarizes the effects of LCD on all four measures of CVD risk. The LCD significantly lowered the Reynolds risk score (RRS), systolic blood pressure ({SBP}) and serum total cholesterol (TBC) but it did not have a statistically significant effect on good cholesterol (HDL). The promising result on the improvement of Reynolds risk score, systolic blood pressure and serum total cholesterol suggests that it may be a reasonable approach, particularly if an individual hopes to avoid medication, to offer LCD with appropriate clinical monitoring. 

\subsubsection{Direct effect of LCD on CVD}

We now study the \emph{natural direct effect} (see, Pearl \cite{pearl2001}) of LCD on CVD risk in the context of the following hypothetical question: 
\begin{quote}
\emph{``For an individual of non-LCD taker, how would LCD affect the risk of CVD?"}
\end{quote}
We are asking what would happen if the treatment, LCD, were to change, but that change did not affect the distribution of the mediator, T2D. 
In that case, the  change in treatment would be propagated only along the direct path  LCD $\to$ CVD in Figure \ref{fig:dag_lcd_cvd2}. 
We argue that the analysis in this situation should control for  gender, age, and T2D, and a look at Figure \ref{fig:dag_lcd_cvd2} give an explanation \cite{pearl2009causality}.  To disable all but the direct path, we need to stratify by T2D. This closes the indirect path
LCD $\to$ T2D $\to$ CVD. But in so doing, it opens two paths LCD $\to$ T2D $\leftarrow$ (Gender, Age) $\to$ CVD, and LCD $\to$ BMI $\to$ T2D $\leftarrow$ (Gender, Age) $\to$ CVD. If we control for (Gender, Age) as well, we close these two paths, and therefore any correlation remaining must be due to the direct path LCD $\to$ CVD. We refer readers to Pearl \cite{pearl2009causality} for an introduction to mediation analysis based on causal diagram. 

To quantify the expected change in CVD if LCD status were changed, we need to control for
calculate
\begin{equation*}
    \mathbb E[\tau_3(\text{Gender}_i,\text{Age}_i, \text{T2D})],
\end{equation*}
where the potential outcome $\tau_3$ is defined as
\begin{equation*}
\begin{aligned}
    \tau_3(\text{Gender}_i,\text{Age}_i, \text{T2D}_i) &= \mathbb E[\text{CVD}_i(\text{LCD}=1)|\text{Gender}_i,\text{Age}_i,\text{T2D}(\text{LCD}=0)] \\
    &\quad\quad -  \mathbb E[\text{CVD}_i(\text{LCD}=0)|\text{Gender}_i,\text{Age}_i].
\end{aligned}
\end{equation*}
The symbol  T2D(LCD $=0$) is the counterfactual distributions of BMI and T2D given that the status of LCD is $0$. The expectations above are taken over the corresponding interventional (i.e., LCD $=0,1$) and counterfactual (i.e.,  T2D(LCD $=0$)) distributions.  We implement I-Rand, which gives the direct effect for the Reynolds risk score in Table \ref{table:maoflcdoncvd}.
Figure \ref{fig:baratelcd}  summarizes the direct effects of LCD on all four measures of CVD risk. 
The LCD has a significant direct effect on lowering the Reynolds risk score (RRS) and serum total cholesterol (TBC) with the average p-value less than $10\%$. 
We complement the results shown in Figure \ref{fig:baratelcd} with the  distributions of ATEs and p-values of the subsamples
in Appendix \ref{sec:lcdcvd}. 
The direct effect in this example represents a stable causal effect that, different from the total effect, is robust to T2D and any cause of CVD risk that is mediated via T2D. 
This robustness makes the natural direct effect a more actionable concept, and in principle, it can be transported to populations with different physical conditions such as T2D status.

\subsubsection{Indirect effect of LCD on CVD}

To isolate the indirect effect from the direct effect, we need to consider a hypothetical change in the mediator while keeping the treatment constant. In our CVD example, we may ask:
\begin{quote}\emph{``How would the CVD risk of an individual without taking LCD be if his / her T2D status had instead following the T2D distribution of individuals taking LCD?"}
\end{quote}
The answer to this question is the average \emph{natural indirect effect} (Pearl \cite{pearl2001}). It can be written as
\begin{equation*}
    \mathbb E[\tau_4(\text{Gender}_i,\text{Age}_i,\text{T2D})],
\end{equation*}
where the potential outcome $\tau_4$ is defined as
\begin{equation*}
\begin{aligned}
    \tau_4(\text{Gender}_i,\text{Age}_i) &= \mathbb E[\text{CVD}_i(\text{LCD}=0)|\text{Gender}_i,\text{Age}_i,\text{T2D}(\text{LCD}=1)] \\
    &\quad\quad -  \mathbb E[\text{CVD}_i(\text{LCD}=0)|\text{Gender}_i,\text{Age}_i].
\end{aligned}
\end{equation*}
The symbol $\text{T2D}(\text{LCD}=1)$ refers to the counterfactual distribution of T2D had LCD been 1, and the expectations are taken over the corresponding interventional (i.e., $\text{LCD}=0,1$) and counterfactual (i.e., $\text{T2D}(\text{LCD}=1)$) distributions.
Under our assumptions, any changes that occur in an individual's CVD risk are attributed to treatment-induced T2D and not to the treatment (i.e., LCD) itself. 

\begin{figure}[!ht]
    \centering
    \includegraphics[width=0.8\textwidth]{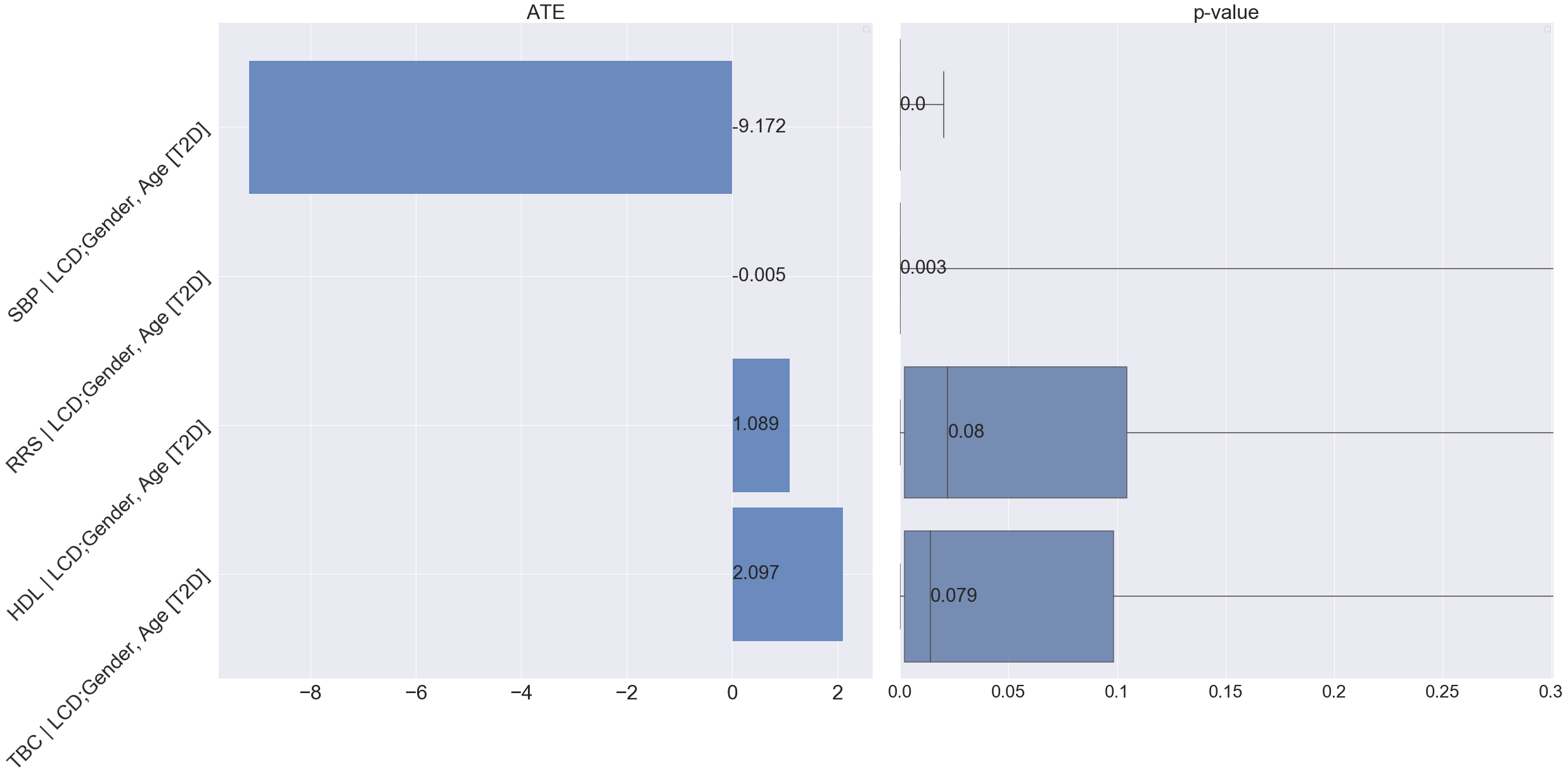}
    \caption{\textbf{Indirect Effect}: Mean ATE bar plot (left) and p-values box plot (right) for the indirect effect of LCD on CVD risk factors with age, gender as confounders and diabetes as a mediator. Each row corresponds to a causal diagram:“Outcome $|$ Treatment; Confounders [Mediator]”.  For example, “SBP $|$ LCD; Gender, Age [T2D]” represents the causal diagram with the systolic blood pressure as the outcome, gender, age, as the confounders, T2D as the mediator, and the LCD as the treatment.}
    \label{fig:nie_barboxlcddiabetes}
\end{figure}

For a  linear model in which there is no interaction between treatment and mediator, the total causal effect can  be decomposed into a sum of direct and indirect contributions (see, e.g., Pearl \cite{pearl2001}):
\begin{equation}
    \label{eqn:decompoftotaleffect}
    \text{total effect} = \text{direct effect} + \text{indirect effect}.
\end{equation}
This decomposition can be applied to each permutation in each subsample.  The estimates are averaged, yielding an estimate of the indirect effect and corresponding distribution of the p-values.
Based on this result, we can assess the indirect effect of LCD on the Reynolds risk score, where the result is provided in Table \ref{table:maoflcdoncvd}.
The negative  sign on the indirect effect indicates that, in addition to its direct effect, the LCD lowered Reynolds risk score through the mediator T2D.
We report the average ATEs and box plots for the distributions of p-values for other CVD risk factors  in Figure
\ref{fig:nie_barboxlcddiabetes}. It shows that the LCD would also have a significant indirect causal effect on other risk factors of CVD, including a reduction in systolic blood pressure ({SBP}) and an improvement in good cholesterol (HDL). We found, however, that the LCD would have a significant indirect effect in the form of an increase in serum total cholesterol (TBC).

\section{Case Study II: Is Obesity A Significant Risk Factor  for T2D and CVD?}
\label{sec:obesity}

\subsection{Causal effect of obesity on T2D}
\label{sec:bmiandt2d}

 \begin{figure}[!ht]
    \centering
    \includegraphics[width=0.4\textwidth]{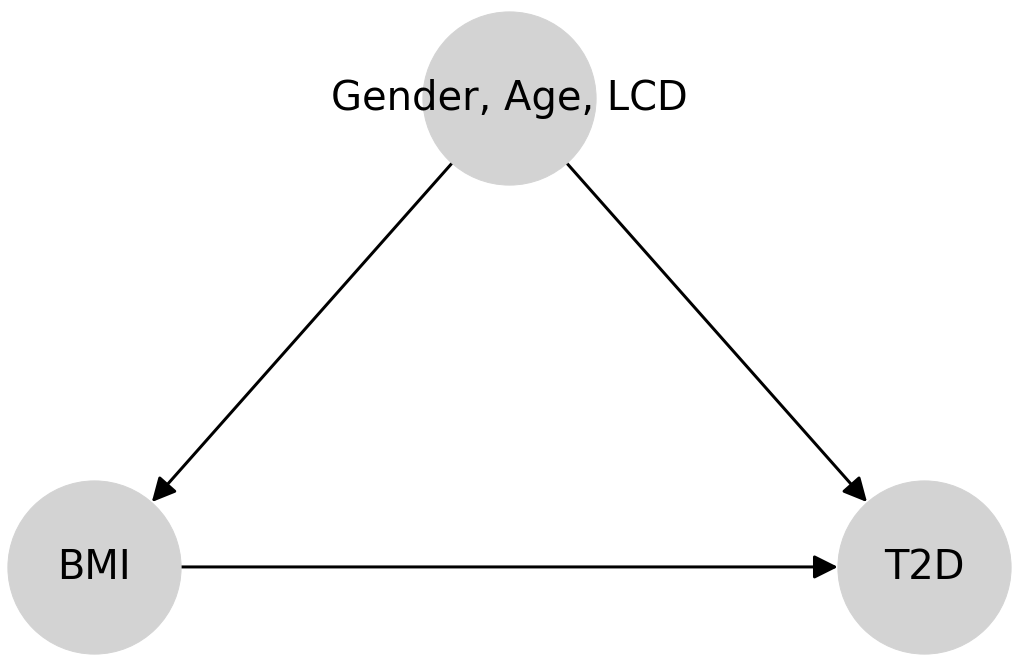}
    \caption{Assumed coarse-grained causal graph for the relationship between BMI and T2D, with  gender and age as confounders.}
    \label{fig:bmit2d}
\end{figure}

Building on the queries in the previous section, we now want to quantify the causal effect of obesity on T2D and CVD \cite{Yoon2006}.
Consider the counterfactual question, 
\begin{quote}
    \emph{``What would be the effect on T2D if an individual changes from normal weight to overweight?"}  
\end{quote}
This question cannot be evaluated with a randomized controlled trial, which would require an experimenter to randomly assign individuals to be either obese or of normal weight.  Instead, we can attempt to estimate the effect of obesity on T2D from
observational data.  We make the following assumptions and a specification of the underlying causal structure. 
First, the  BMI is modeled as a categorical variable in this section: \emph{normal} weight if $\text{BMI}<25$,  \emph{overweight} if $\text{BMI} \in[25,30)$, \emph{obese} if $\text{BMI}\in[30,35)$), and \emph{severely obese} if $\text{BMI}\geq 35$.
In our analysis, we compare consecutive ordinal levels of obesity pairwise. At each time, we denote the higher level of obesity as 1 (treatment) and the lower level of obesity as 0 (control).
Second, similar to the motivating example in Section \ref{sec:motivating}, gender is a binary variable and  age is an ordinal variable, and the medical outcome T2D is an ordinal variable indicating status at time of reporting: non-diabetics,  pre-diabetics, and  diabetics.
Finally, we assume the causal graph shown in Figure \ref{fig:bmit2d}, and motivate it by thinking of the following data-generating process:
(1) BMI affects the risk of T2D;
(2) Gender, age and LCD are unaffected by the BMI level;
(3) Gender, age and LCD affect the risk of T2D and the BMI level. Thus gender, age and LCD are confounders of BMI and T2D.
(4) Causal sufficiency:  there are no hidden confounders. 
Under these assumptions, we can calculate an estimate of the effect of BMI on T2D, by adjusting for the confounders using the model of potential outcomes in Section \ref{sec:potentialoutcome}.

\begin{table}
\caption{Average treatment effect of BMI on T2D: $\mathbb E[\tau_5]$}
\centering
\begin{tabular}{ c c c c}
\toprule
 & Normal weight \textit{vs.} overweight & Overweight \textit{vs.} obese & Obese \textit{vs.} severely obese\\
\midrule
 ATE & 0.477 & 0.316 & 0.14\\
 p-value & 0.002 & 0.011 & 0.196\\
\bottomrule
\end{tabular}
\label{table:ateofbmiont2d}
\end{table}

\begin{figure}[!ht]
    \centering
    \includegraphics[width=\textwidth]{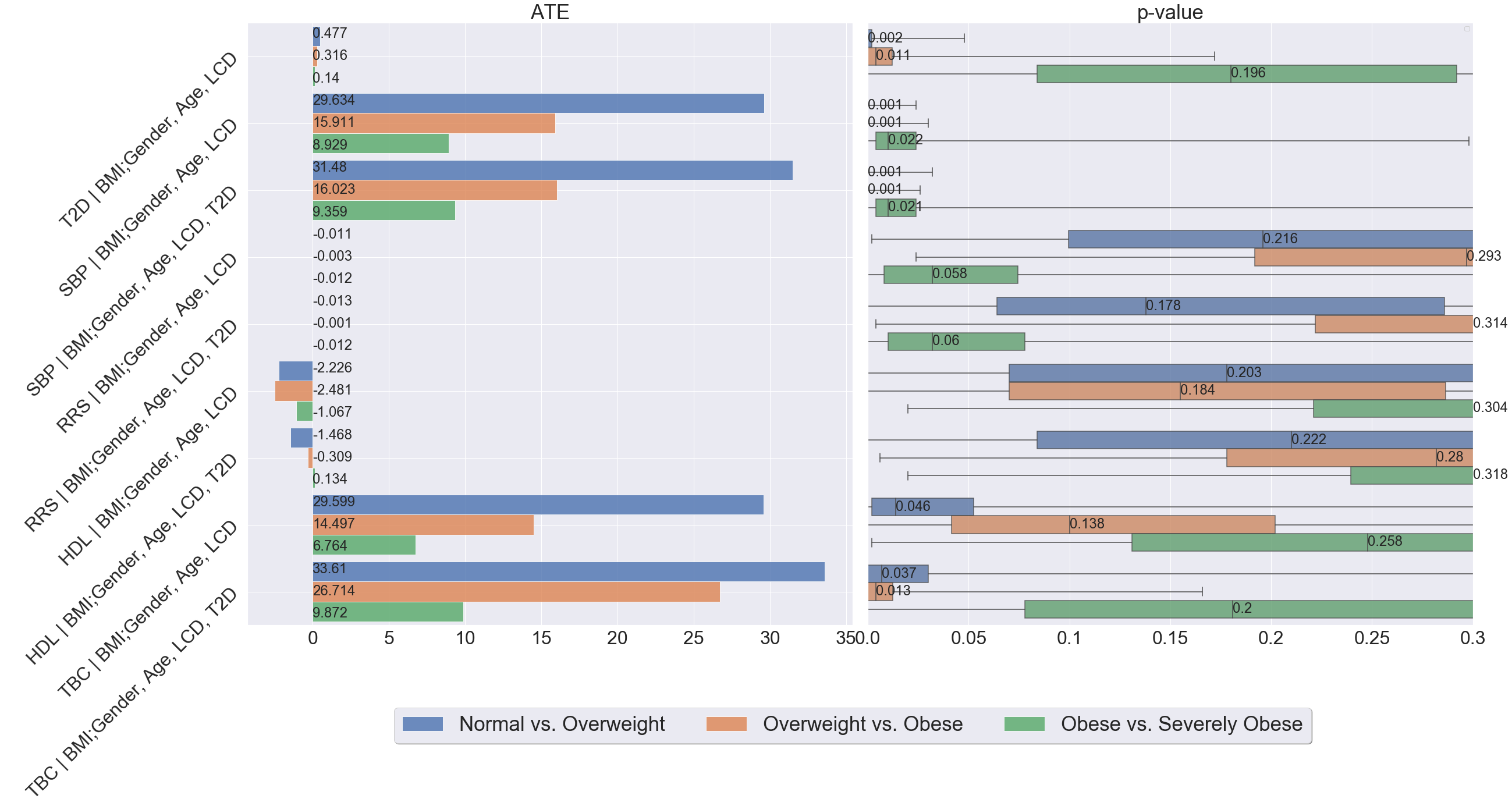}
    \caption{Mean ATE bar plot (left) and p-value box plot (right) for BMI as the treatment. Each row corresponds to a causal diagram: ``\emph{Outcome $|$ Treatment; Confounders}". For example, ``\emph{SBP $|$ BMI; Gender, Age, T2D}" represents the the causal diagram with the systolic blood pressure as the outcome, and the gender, age, T2D as the confounders, and the BMI as the treatment which takes three pairwise comparisons: normal weight vs. overweight (green), overweight vs. obesity (orange), obesity vs. severe obesity (blue).}
    \label{fig:baratebmi}
\end{figure}

According to the causal graph in Figure \ref{fig:bmit2d}, BMI takes the role of a treatment that affects the outcome T2D. To quantify the expected change in T2D if BMI were changed, we need to calculate
\begin{equation*}
    \mathbb E[\tau_5(\text{Gender}_i,\text{Age}_i,\text{LCD}_i)],
\end{equation*}
where the potential outcome $\tau_5$ is defined as
\begin{equation*}
    \tau_5(\text{Gender}_i,\text{Age}_i,\text{LCD}_i) = \mathbb E[\text{T2D}_i(\text{BMI}=1)|\text{Gender}_i,\text{Age}_i,\text{LCD}_i]  -  \mathbb E[\text{T2D}_i(\text{BMI}=0)|\text{Gender}_i,\text{Age}_i,\text{LCD}_i].
\end{equation*}
By I-Rand in Algorithm \ref{alg:i-rand}, we obtain the mean of ATEs $\mathbb E[\tau_1]$ over $500$ subsamples and the mean p-value (from the permutation tests)  as follows (see, also Figure \ref{fig:baratebmi}) for all three pairwise differences: (1) changing from normal weight to overweight; (2) changing from overweight to obese; (3) changing from obese to severely obese. 

We summarize the results in Table \ref{table:ateofbmiont2d}, which suggests that the difference of T2D constitutes a causal effect, and changing BMI level from a lower level to a higher level would lead to an increased risk of T2D, where the results are subject to our modelling assumptions. 
We note that under the null hypothesis of no causal effect,
the p-values follow a uniform distribution on $(0,1)$ given sufficiently many subsamples. However, the box plot of p-values in Figure \ref{fig:baratebmi}, corresponding to  the causal graph in Figure \ref{fig:bmit2d}, shows p-values are concentrated at the origin, which indicates a strong evidence for the alternative hypothesis. 
In particular, the causal effect of the treatment (normal weight vs. overweight) with p-value $0.002$ is significant under the Bonferroni's false discovery control at the $0.01$ level. 
The detailed distributions of ATEs and p-values are provided in Appendix \ref{sec:suppnum}, which confirms the consistency of these results across subsamples.

\subsection{Mediation analysis for the effect of obesity on CVD}
\label{sec:obsandcvd}


We now seek to understand the role of T2D as a mediator of the effect of  obesity on CVD risk. As discussed in Section \ref{sec:medationlcdoncvd}, this mediation analysis is particularly relevant from the perspective of an individual with both T2D and CVD.
We study four well-known risk factors of CVD: systolic blood pressure, serum cholesterol level, high-density lipoprotein, and Reynolds risk score; see, Ridker et al. \cite{ridker2007}. 


\subsubsection{Causal graph of T2D as a mediator} 

\begin{figure}[!ht]
    \centering
    \includegraphics[scale=0.2]{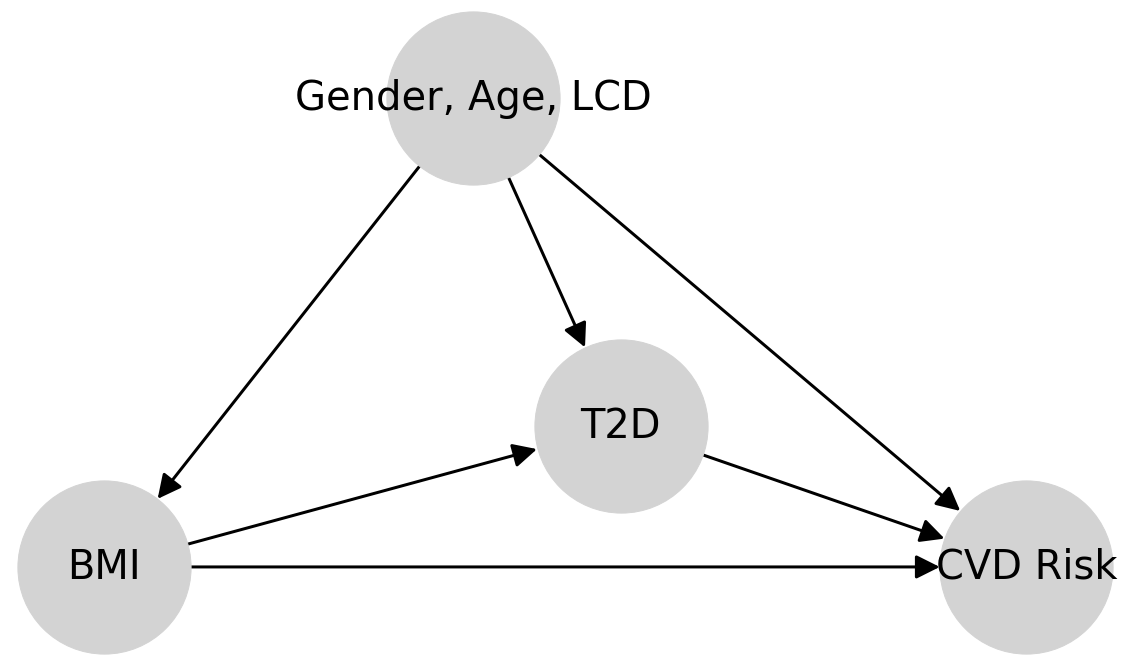}
    \caption{Assumed coarse-grained causal graph for the relationship between BMI, T2D, and the outcome CVD. Within this view, T2D acts as a \emph{mediator} of the effect of BMI on CVD,  with the gender, age and LCD as confounders.}
    \label{fig:dag_bmi_cvd2}
\end{figure}
We assume the causal graph in Figure \ref{fig:dag_bmi_cvd2}, and  motivate Figure \ref{fig:dag_bmi_cvd2} with the following data-generating process:
(1) Choose a BMI level at random;
(2) Given a selected BMI level, sample an individual with a T2D status;
(3) Conditional on the choice of BMI level and T2D status, sample the medical outcome within a given CVD risk factor.
The details are as follows.
First, the arrow BMI $\to$ T2D encodes that the distribution of T2D depends on BMI level. This dependence was quantified in Section \ref{sec:bmiandt2d}. 
Second, the arrow T2D $\to$ CVD reflects the established knowledge in nutrition science that T2D influences  CVD risk \cite{Mar2014,benjamin2018}. 
Finally,  since our model assumes causal sufficiency, and in particular, that  T2D is the only mediator in the effect of BMI on CVD risk,  the arrow BMI $\to$ CVD represents obesity-specific influences on CVD risk.  
In addition to the causal graph in Figure \ref{fig:dag_bmi_cvd2}, we assume there are no hidden confounders. 
Given these assumptions, we see that BMI causally influences CVD risk along two different paths:
 a  path BMI $\to$ CVD, giving rise to a \emph{direct effect}, and a path BMI $\to$ T2D $\to$ CVD mediated by T2D, giving rise to an \emph{indirect effect}.
Note that the direct effect of BMI on CVD is likely mediated by additional variables that are subsumed in BMI $\to$ CVD Risk. We discuss this point further in Section \ref{sec:discussion}.
In mediation analysis, the goal is to quantify direct and indirect effects. 
We start with the total effect and then formulate the direct and indirect effects by allowing the treatment to propagate along one path while controlling the other path.

\subsubsection{Total effect of BMI on CVD}

Given the causal assumptions in the previous section, the first measure  of interest is the total causal effect of obesity on CVD, i.e., the answer to the following question:
\begin{quote}
\emph{``What would be the effect on CVD if an individual changes from normal weight to overweight?"}
\end{quote}
As we did in Section \ref{sec:bmiandt2d}, we  formulate the answer using the ATE:
\begin{equation*}
    \mathbb E[\tau_6(\text{Gender}_i,\text{Age}_i,\text{LCD}_i)],
\end{equation*}
where the potential outcome $\tau_6$ is defined as
\begin{equation*}
    \tau_6(\text{Gender}_i,\text{Age}_i,\text{LCD}_i) = \mathbb E[\text{CVD}_i(\text{BMI}=1)|\text{Gender}_i,\text{Age}_i,\text{LCD}_i]  -  \mathbb E[\text{CVD}_i(\text{BMI}=0)|\text{Gender}_i,\text{Age}_i,\text{LCD}_i].
\end{equation*}
We now give a detailed result for one of the CVD risk factors, namely the systolic blood pressure,
where the description and the summary statistics are deferred to Appendix \ref{appendixA}.
It is known that increasing systolic blood pressure  significantly increases the risk of CVD (e.g., Bundy et al. \cite{Bundy2017}). 
By the proposed I-Rand with $500$ subsamples and the corresponding permutation test for each subsample, we obtain the mean of ATEs $\mathbb E[\tau_6]$ given in Table \ref{table:maofbmioncvd}. 
\begin{table}
\caption{Mediation analysis for the effect of BMI on SBP}
\centering
\begin{tabular}{ c c c c}
\toprule
 & Normal weight \textit{vs.} overweight & Overweight \textit{vs.} obese & Obese \textit{vs.} severely obese\\
\midrule
Total effect $\mathbb E[\tau_6]$ (p-value) & 29.634 (0.001) & 15.911 (0.001) & 8.929 (0.022)\\
Direct effect $\mathbb E[\tau_7]$ (p-value) & 31.48 (0.001) & 16.023 (0.001) & 9.359 (0.021)\\
Indirect effect $\mathbb E[\tau_8]$ (p-value) &-1.846 (0.112) & -0.112 (0.146)  & -0.43 (0.231)\\
\bottomrule
\end{tabular}
\label{table:maofbmioncvd}
\end{table}
The results show that  an individual changing from normal weight to overweight would significantly lead to an increase in  systolic blood pressure.  
In contrast,  the box plot of p-values in Figure \ref{fig:baratebmi} indicates only weak evidence that changing BMI would have a causal effect on other risk factors of CVD including serum total cholesterol (TBC), high-density lipoprotein (HDL), and Reynolds risk score (RSS). 
The observation is also supported by distributions of ATEs and p-values in Appendix \ref{sec:suppnum}. 
The  failure
to reject the null hypothesis may also be due to unobserved confounders such as  genetic information, smoking, and stress levels.

\subsubsection{Direct effect of BMI on CVD}

We now study the \emph{natural direct effect} (see, Pearl \cite{pearl2001}) of obesity on CVD risk in the context of the following hypothetical question: 
\begin{quote}
\emph{``For an individual of normal weight, how would a weight gain  affect the risk of 
 CVD?"}
 \end{quote}
We are asking what would happen if the treatment, BMI, were to change, but that change did not affect the distribution of the mediator, T2D. 
In that case, the  change in treatment would be propagated only along the direct path  BMI $\to$ CVD in Figure \ref{fig:dag_bmi_cvd2}. 
To disable all but the direct path, we need to stratify by T2D. This closes the indirect path
BMI $\to$ T2D $\to$ CVD. But in so doing, it opens the path BMI $\to$ T2D $\leftarrow$ (Gender, Age, LCD) $\to$ CVD since T2D is a collider in Figure \ref{fig:dag_bmi_cvd2}. If we control for (Gender, Age, LCD) as well, we close the direct path, and therefore any correlation remaining must be due to the direct path BMI $\to$ CVD.

To quantify the expected change in T2D if BMI were changed, we need to calculate
\begin{equation*}
    \mathbb E[\tau_7(\text{Gender}_i,\text{Age}_i,\text{LCD}_i,\text{T2D})],
\end{equation*}
and where the potential outcome $\tau_7$ is defined as follows:
\begin{equation*}
\begin{aligned}
    \tau_7(\text{Gender}_i,\text{Age}_i,\text{LCD}_i,\text{T2D}_i) &= \mathbb E[\text{CVD}_i(\text{BMI}=1)|\text{Gender}_i,\text{Age}_i,\text{LCD}_i, \text{T2D}(\text{BMI}=0)]  \\
    &\quad\quad -  \mathbb E[\text{CVD}_i(\text{BMI}=0)|\text{Gender}_i,\text{Age}_i,\text{LCD}_i].
\end{aligned}
\end{equation*}
The symbol $\text{T2D}(\text{BMI}=0)$ refers to the counterfactual distribution of T2D given that the value of BMI is 0, and the expectations are taken over the corresponding interventional (i.e., $\text{BMI}=0,1$) and counterfactual (i.e., $\text{T2D}(\text{BMI}=0)$) distributions. 
Hence, $\tau_7$ defines the influence that is not mediated by T2D in the sense that it quantifies the sensitivity of the CVD to changes in BMI while T2D 
is held fixed, as illustrated in Figure \ref{fig:dag_bmi_cvd2}.
By I-Rand algorithm, we obtain mean ATEs $\mathbb E[\tau_7]$ with $500$ subsamples and the mean p-value of permutation tests for systolic blood pressure in Table \ref{table:maofbmioncvd}. See, also Figure \ref{fig:baratebmi} for other CVD risk factors.
In addition to the summary statistics shown above, we provide the distributions of ATEs and p-values of the subsampling in Appendix \ref{sec:obsitycvd}.
We find, for example, that a change from normal weight to overweight would lead to a increase in systolic blood pressure of $31.48$ mmHg on average (see Appendix \ref{appendixA} for summary statistics of systolic blood pressure). 
 The direct effect in this example represents a stable biological relationship that, different from the total effect, is robust to T2D and any cause of high systolic blood pressure that is mediated via T2D. 

\subsubsection{Indirect effect of BMI on CVD}

We conclude this section by studying the indirect effect in the context that 
\begin{quote}
\emph{``How would the CVD risk of a normal weight individual be  if his / her T2D status had instead following the T2D distribution of overweight individuals?"}
\end{quote}
The answer is formulated by
\begin{equation*}
    \mathbb E[\tau_8(\text{Gender}_i,\text{Age}_i,\text{LCD}_i,,\text{T2D})],
\end{equation*}
where the potential outcome $\tau_8$ is defined
\begin{equation*}
\begin{aligned}
    \tau_8(\text{Gender}_i,\text{Age}_i) &= \mathbb E[\text{CVD}_i(\text{BMI}=0)|\text{Gender}_i,\text{Age}_i,,\text{LCD}_i \text{T2D}(\text{BMI}=1)]  \\
    &\quad\quad -  \mathbb E[\text{CVD}_i(\text{BMI}=0)|\text{Gender}_i,\text{Age}_i,\text{LCD}_i].
\end{aligned}
\end{equation*}
Under our assumptions,  any changes that occur in an individual's CVD risk are attributed to BMI-induced T2D and not to the BMI itself. 
The indirect effect of the treatment is 
the change of CVD risk obtained by 
keeping the BMI of each individual fixed and setting the distribution of T2D to the level obtained under treatment.

\begin{figure}[!ht]
    \centering
    \includegraphics[width=0.9\textwidth]{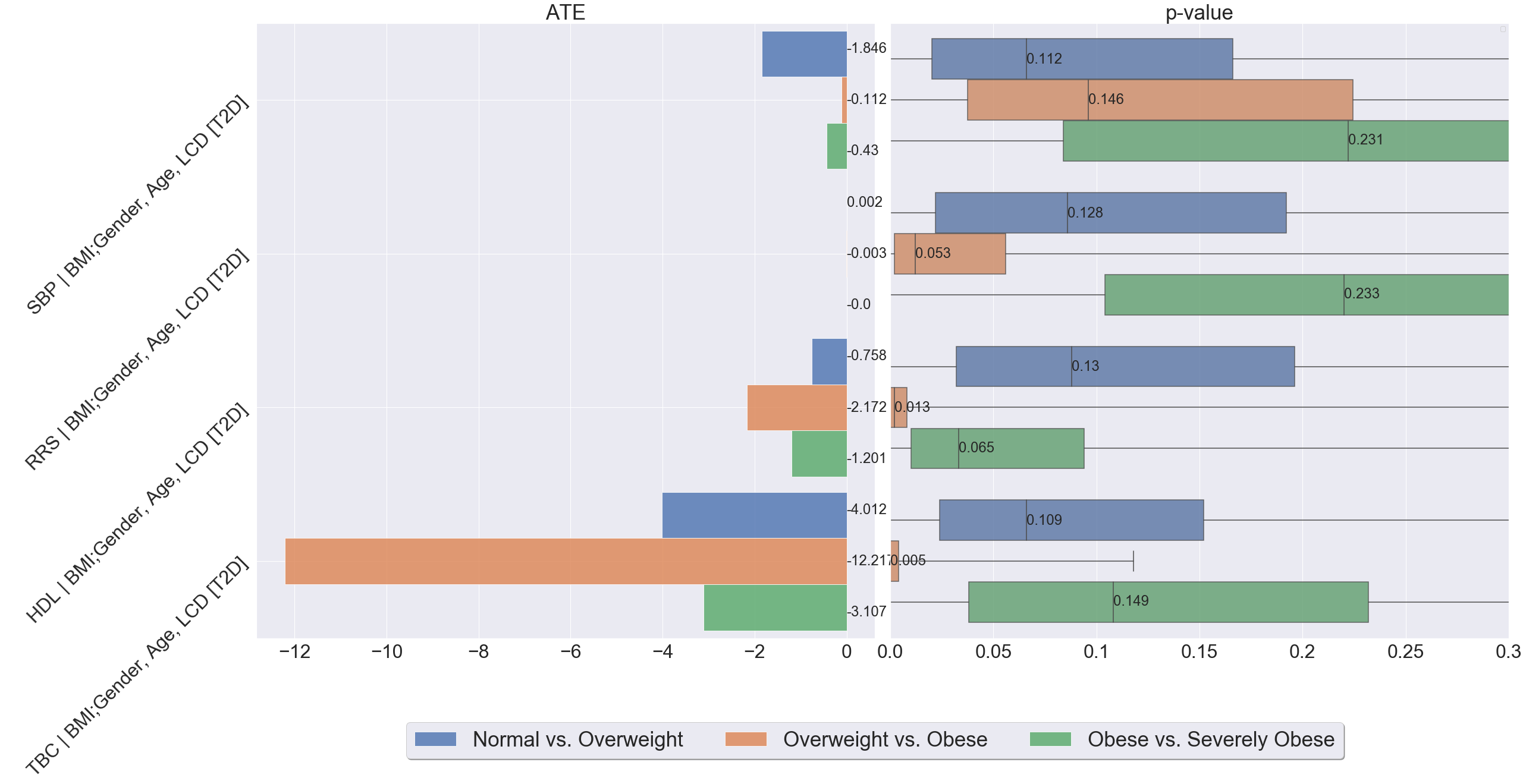}
    \caption{\textbf{Indirect Effect}: Mean ATE bar plot (left) and p-values box plot (right) for the indirect effect of BMI on CVD risk factors with age and gender as confounders and T2D as a mediator. Each row corresponds to a causal diagram: “Outcome $|$ Treatment; Confounders [Mediator]”.  For example, “SBP $|$ BMI; Gender, Age, [T2D]” represents the causal diagram with the systolic blood pressure as the outcome, gender and age as the confounders, T2D as the mediator, and BMI as the treatment.}
    \label{fig:nie_barboxbmidiabetes}
\end{figure}

Consider a linear model in which there is no interaction between treatment and mediator. This yields the decomposition (\ref{eqn:decompoftotaleffect}) and the indirect effect of BMI on the systolic blood pressure given in Table \ref{table:maofbmioncvd}. 
We report the average ATEs and box plots for the distributions of p-values for other CVD risk factors in Figure \ref{fig:nie_barboxbmidiabetes} .
We find that changing only the distribution of T2D that results from an increase in BMI from normal weight to overweight would lead to a decrease in systolic blood pressure of about $1.848$ mmHg on average. Notably, the sign of this indirect effect is opposite to the sign of the corresponding direct effect, which suggests that indirect and direct effects tend to offset one another. 
There are several possible explanations for this. For example, the offset may be due to missing BMI data,
which results in selection bias.
Further discussion of selection bias is  in Section \ref{sec:discussion}. Another possible explanation is the obesity paradox given the comorbidity conditions (see, e.g., Uretsky et al. \cite{uretsky2007} and Lavie et al. \cite{lavie2009}) that overweight people may have a better prognosis,
possibly because of the medication or overweight individuals having lower systemic vascular resistance  compared to leaner hypertensive individuals.

\section{Discussion on  Assumptions and Models}
\label{sec:discussion}

We assume the causal relationships between variables of demographics, obesity, T2D, and CVD to be captured by causal graphs in the previous sections,
which correspond to different nutrition-related questions. These causal graphs constitute a coarse-grained view, which neglects
many 
potentially important risk factors.
A strength of this coarse-grained approach is that it allows for quantitative reasoning about different causal effects including total, direct, and indirect effects in situations where the data do not allow a more fine-grained analysis. 
In the following, we discuss assumptions and limitations of our approach and point out some future directions.

\subsection{Selection bias}

The data we considered concerns only those patients who are from the Norwood general practice surgery in England and has opted to follow LCD by 2019 \cite{unwin2020insights}.
We can introduce an additional variable $V$ with $V=1$ meaning that an individual who is from the Norwood general practice surgery and follows LCD by 2019 and $V=0$ otherwise. In that case, our analysis is always conditioned on $V=1$. If the individual who follows LCD is randomly sampled from the population of Norwood general practice surgery with 9,800 patients, the implicit conditioning on $V=1$ would not introduce bias to an inference for the larger population. 
However, samples are generally not collected randomly. In particular, age and health conditions are causal factors on the participation in the LCD program, i.e., age $\to V$ and health condition $\to V$ and  through self-selection. Moreover, due to the possible speciality and reputation of the LCD program, T2D$\to V$, CVD$\to V$. Finally, there may be complex interactions between office visit and T2D or CVD, where the process involves the feedback $V \to$ T2D and $V\to$ CVD.
The fact that we consider only individuals who participate in the LCD program while the visit itself depends on multiple other factors inevitably leads to the problem of \emph{selection bias}.
Several approaches have been developed to decrease this bias under certain conditions; see, e.g., Bareinboim and Tian \cite{Bareinboim2015}, Bareinboim and Pearl \cite{Bareinboim2016}. 

\subsection{Unobserved confounders}
{\color{black} An important assumption upon which relies our estimation of the causal effect, is the absence of hidden confounders, i.e., we assume that gender and age are the only confounders\footnote{One can argue that gender and age are insufficient confounders for the analysis of CVD risk factors, however, the current data at hand only allows for these.}. In particular, it is the basis of our estimates of the direct and indirect effects.}
It may be possible to relax the absence of hidden confounders depending on the availability of experimental data. See Pearl \cite{pearl2001} for further discussion.

\subsection{Additional mediators}

In our coarse-grained view, the arrows possibly subsume many other potentially important risk factors within the causal paths.
For example, the strength of the effect BMI $\to$ T2D in Section \ref{sec:motivating} is estimated without consideration of mediators.

\subsection{Model selection}
In this section, we compare the proposed I-Rand with the  method of difference-in-differences (DD; e.g., Angrist and Pischke \cite{Angrist2008}), which studies the impact of the differential effect of a treatment on a ``treatment group" versus a ``control group". Then we discuss some generalizations of the models used for analysis in the previous sections.
The following analysis explores the impact of difference in treatment on difference in outcome (DD). For a given variable, we calculate the difference as the value on the second visit minus the value on the first visit.  In our analysis, we set  confounders  (e.g., age and gender) to the values  recorded at the first visits. We note two main differences between I-Rand and DD.
First, when the LCD is the treatment, all individuals are LCD-takers and there is no control group. The  I-Rand creates a control group by subsampling, while difference-in-differences relies on the null hypothesis of ``no effect.''
Second, when BMI is the treatment, I-Rand subsamples one of the two observations for each individual to avoid  two types of unintended treatments. In DD, such subsampling is unnecessary since we have one observation for each individual. We perform two experiments: a decrease in BMI (i.e., $\Delta \text{BMI} < 0$), or a change in BMI in excess of a threshold  (e.g., $\Delta\text{BMI}$  $<\text{median of }|\Delta\text{BMI}_i|$). The latter choice of treatment splits the data into two equally-sized subgroups of treatment and control, and it is more robust than the first choice of treatment since  the BMI of almost all individuals decreased between visits. For BMI with a median threshold, the causal effect for individual $i$ has the usual estimation formula, i.e., ATE $= \mathbb{E}|\tau(X_i)]$, where $\tau(X_i) = \mathbb{E}[Y_i(1) | X_i] - \mathbb{E}[Y_i(0) | X_i]$. In the case of LCD and decrease of BMI, the causal effect reduces to $\tau(X_i) = \mathbb{E}[Y_i(1) | X_i]$. Note that the design of the experiment however, breaches the non-zero probability of receiving treatment assumption, i.e.,  $0 < P(T = 1|X) < 1$, which is required by the causal effect estimation. As a matter of fact, all individuals are treatment-takers between the two observation dates. Hence, we add a hypothetical control group that does not take the treatment and has a $0$ valued outcome. To estimate the causal effect for the latter that is applicable to permutation analysis, we implement  Algorithm \ref{alg:did} for difference-in-differences analysis.
\begin{algorithm}
\caption{ \normalsize{Difference-in-difference with data re-organization of two-point time-series without control group }}\label{alg:did}
\begin{algorithmic}[1]
\State  \normalsize{\textbf{Input:} Observed data of confounders, treatment and outcome $(X_{i, t}, T_{i, t}, Y_{i, t})$ where $i$ is the individual's ID and $t$ is the state.}
\State Create the ``treatment'' difference-in-differences matrix with $T_{i, 1} = T_{i, t=1} - T_{i, t=0}$, $Y_{i, 1} = Y_{i, t=1} - Y_{i, t=0}$, $X_{i, 1} = X_{i, t=0}$. Its ``control'' image with the following attributes:  $T_{i, 0} = 0$, $Y_{i, t=0} = 0$, $X_{i, 0} = X_{i, t=0}$. Form the difference-in-differences matrix as the concatenation of the two. 
\State Calculate ATE by the matching method over the concatenated matrix. Here,  the estimation of the ATE leads to $\mathbb{E}[Y_i(1) | X_i] - \mathbf{0}$ since the control set consists of only $0$ valued outcomes. In the permutation analysis on the other hand, the estimation results in a difference of the two usual quantities.
\State  Perform the permutation analysis: \textbf{for} $s=1,2,\ldots, S$ \textbf{do}
\State \quad\quad Sample a binary vector of length $N$, where the index is the individual's ID and the value is the state (sampling without replacement). Selecting the corresponding subsample $s$ as the shuffled vector of treatment. Note that the shuffling performed is equivalent to assigning each individual the treatment with probability 1/2 ($N$ independent Bernoulli variables with parameter $p=1/2$). This is a strong assumption that needs to be supported by data. However, this choice is consistent with our I-Rand algorithm and can further be adjusted to a better choice of $p$;
\State \quad\quad Calculate $ATE^{(s)}$ for the shuffle $s$ of the treatment.
\State \textbf{end for}
\State Calculate the p-value$= \frac{1}{S}\sum_{s=1}^{S}\mathbbm{1}_{ATE^{(s)} > (\text{resp. } <) ATE}$  for the one-tailed test for the null hypothesis of no treatment effect.
\State \textbf{Output:} ATE and p-value.
\end{algorithmic}
\end{algorithm}

\begin{figure}[!ht]
    \centering
    \includegraphics[width=0.5\textwidth]{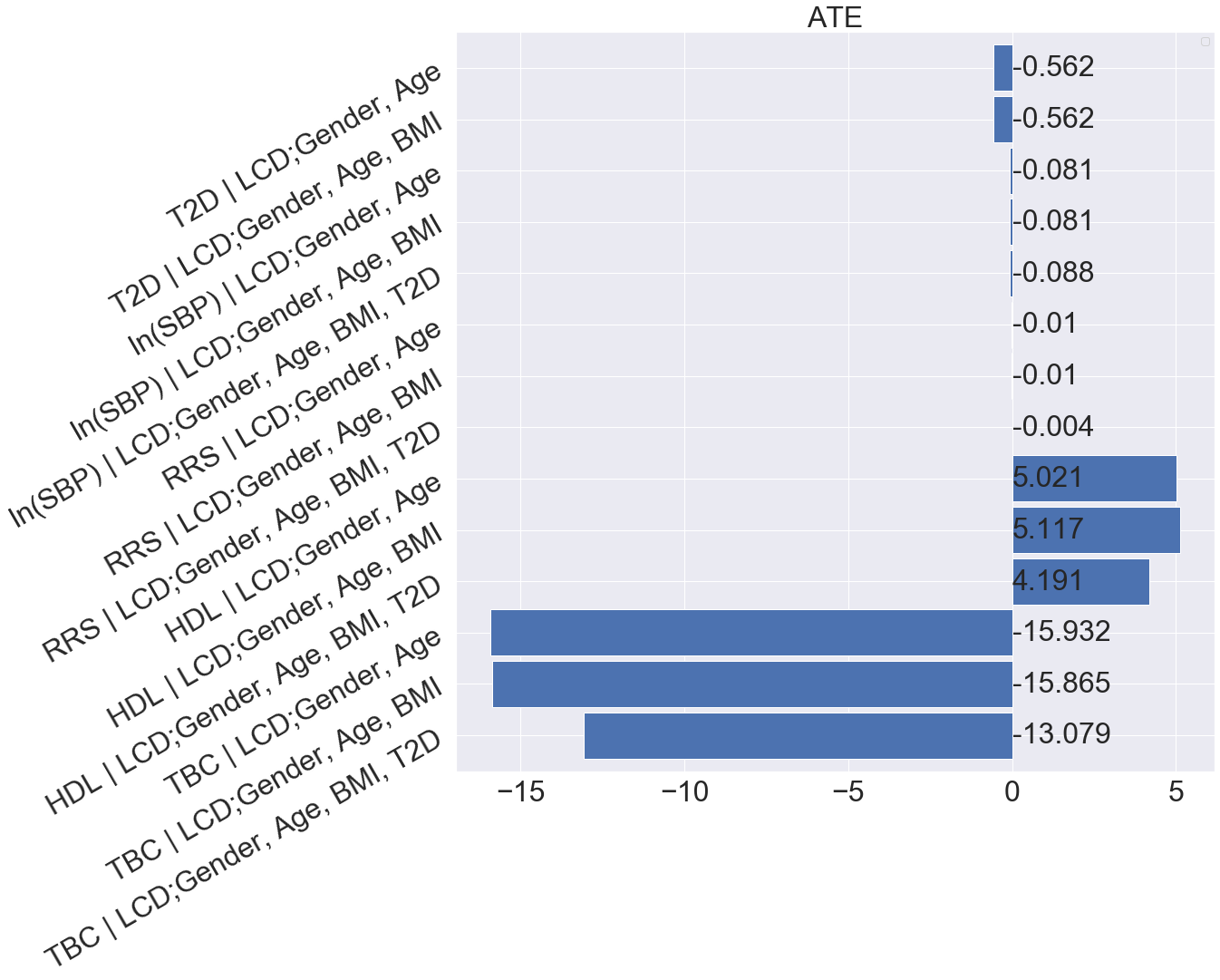}
    \caption{Bar plot of ATE for LCD as the treatment for the difference-in-differences analysis without threshold (we omit the p-values as they are all under a 1\% significance level). Each row corresponds to a causal diagram:“Outcome $|$ Treatment; Confounders”.  For example, “SBP|LCD; Gender, Age, BMI” represents the causal diagram with the systolic blood pressure as the outcome, and gender, age, BMI as the confounders, and LCD as the treatment.}
    \label{fig:barDiD_CLD}
\end{figure}

\begin{figure}[!ht]
    \centering
    \includegraphics[width=0.5\textwidth]{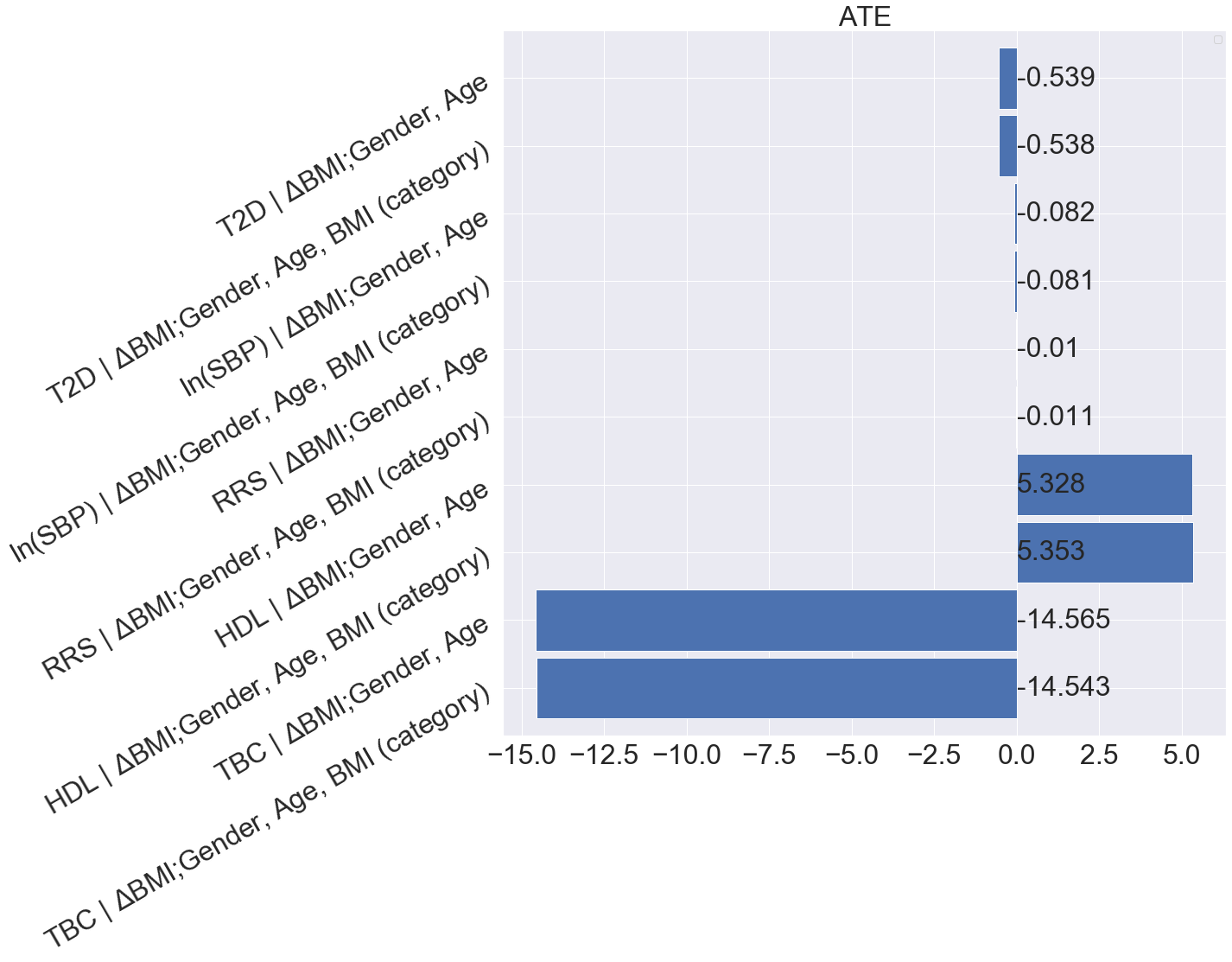}
    \caption{Bar plot of ATE for BMI as the treatment for difference-in-differences analysis without threshold (we omit the p-values as they are all under a 1\% significance level). Each row corresponds to a causal diagram:“Outcome $|$ Treatment; Confounders”.  For example, “SBP $|$ BMI; Gender, Age” represents the causal diagram with the systolic blood pressure as outcome, gender and age as confounders, and BMI as  treatment.}
    \label{fig:barDiD_BMI}
\end{figure}

\begin{figure}[!ht]
    \centering
    \includegraphics[width=0.8\textwidth]{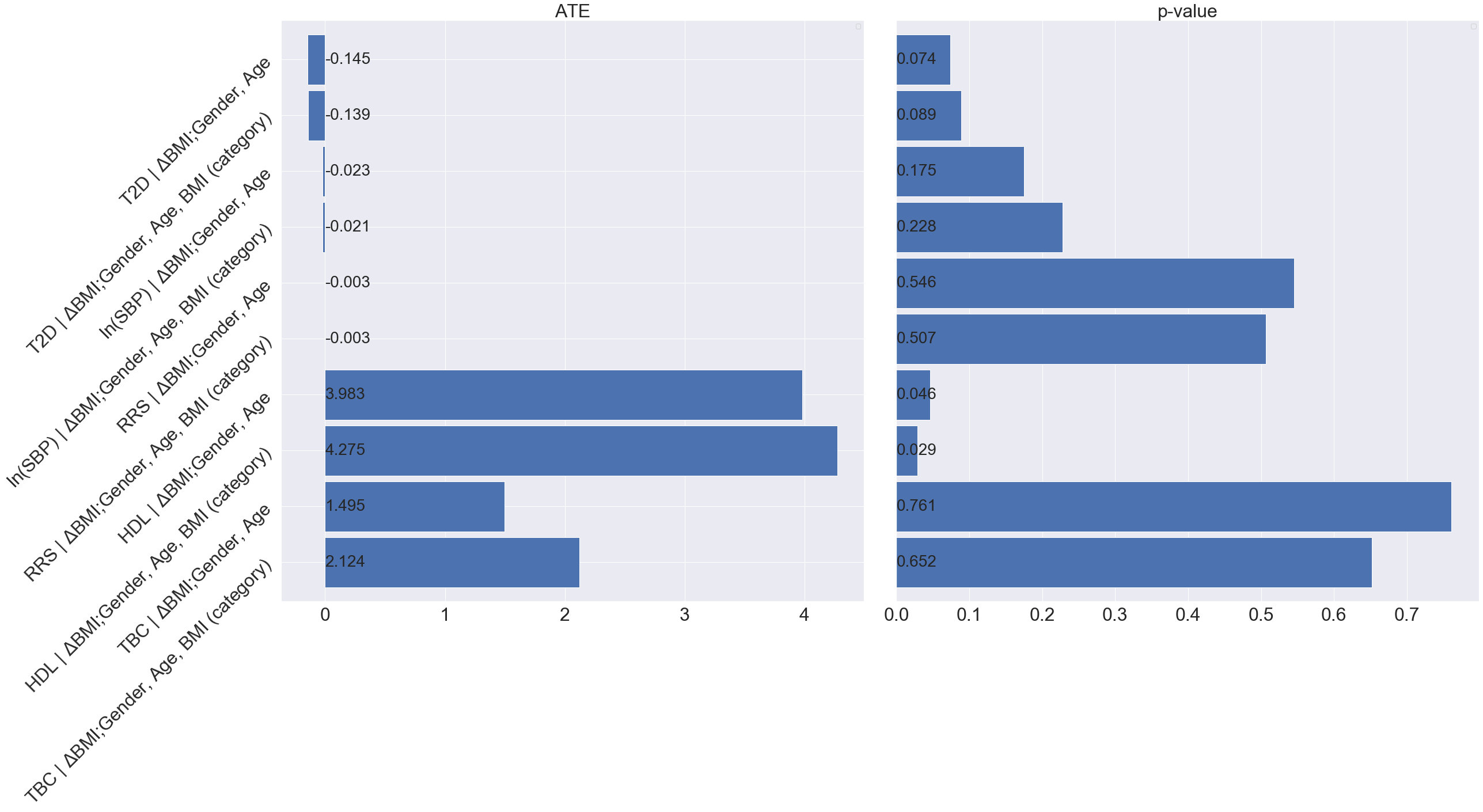}
    \caption{Bar plot of ATE (left) and p-value (right) for BMI as the treatment for the difference-in-differences analysis with threshold (median).  Each row corresponds to a causal diagram:“Outcome $|$ Treatment; Confounders”.  For example, “SBP $|$ BMI; Gender, Age” represents the causal diagram with the systolic blood pressure as the outcome, and gender, age as the confounders, and BMI as the treatment.}
    \label{fig:barDiD_BMI_median}
\end{figure}

We summarize the results in Figures \ref{fig:barDiD_CLD}, \ref{fig:barDiD_BMI}, and \ref{fig:barDiD_BMI_median}.
For a change in diet (i.e., $\Delta \text{LCD}=1$), we find that LCD diet significantly impacts the change in T2D status and CVD risk factors.
The same applies to the first choice of treatment for BMI (i.e., $\Delta$ BMI $< 0$). 
For the second choice of treatment for BMI (i.e., $\Delta$ BMI $< \text{Threshold}$), we find that a change in BMI has a significant causal effect on a change in T2D even when controlling for the BMI categories. Moreover,  a decrease in BMI leads to an increase in HDL (ATE$=3.983$, p-value$<4.6\%$ without controlling for BMI categories and ATE$=4.275$, p-value$<2.9\%$ when controlling for BMI categories).

The limitation of the difference-in-differences in our dataset is that we do not have enough data for longitudinal analysis. As a result, variance across samples (noise) could be much larger than the variance within samples (signal).
On the other hand, the results based on the difference-in-differences method with only two time points are always subject to biases (e.g., Raudenbush \cite{Raudenbush2001}).

There are different ways to generalize our linear models to nonlinear models. For example, it could be of interest to  ask whether or not BMI above a certain threshold has a causal effect on CVD or T2D.
Moreover, the linear models we used in this article cannot represent interactions among variables. It could also be of interest to assess the direct and indirect effects allowing for interactions between treatments and mediators \cite{pearl2001}.



\section*{Acknowledgements}
The authors gratefully acknowledge support from
the Consortium for Data Analytics in Risk, Swiss Re Institute, and NNEdPro Global Centre for Nutrition and Health. 
We are very grateful to Bob Anderson,  Nate Jensen, David Unwin, and participants of the Risk Seminar at UC-Berkeley for  insightful comments on this work.


\bibliography{causal-nutrition}

\appendix

\section{Matching Methods}
\label{sec:matching}
\subsection{Distance measure based on propensity score}

We need to determine which confounders to include for matching and to combine those variables into one measure. 
Under the strong ignorability assumption, it is necessary to include all variables known to be related to both treatment assignment and the outcome in the matching procedure \cite{heckman1998, rubin1996}.
There is little cost to including variables that are not associated with treatment assignment. However, excluding a potentially important confounder can yield a large bias.
In the other direction, variables such as colliders and mediators that may have been affected by the treatment should be excluding from the matching process, and should be used instead in the analysis model for outcomes (see, Greenland \cite{greenland2003}).

The \emph{propensity score}, a popular measure to combine confounders, is defined for each individual $i$  as the probability of receiving the treatment, given the observed confounders \cite{rosenbaum1983central}:
\begin{equation*}
e_i(X_i) = \mathbb P(T_i=1|X_i).
\end{equation*}
The propensity score has two well-known properties. First,
a propensity score is a balancing score in the sense that at any level of the propensity score, the distributions of the confounders defining the propensity score in the treated and control groups are the same.  Second, the treatment assignment is ignorable given the propensity score if treatment assignment is
ignorable given the confounders. Hence, it is reasonable to 
match individuals on the basis of  propensity score rather
than the vector of multivariate confounders.

These properties imply that  the difference in means for the outcomes between treated and control
individuals with a particular propensity score value is
an unbiased estimate of the treatment effect at that
propensity score value. 
The distance between individuals $i$ and $j$ through the propensity score is
$D_{ij} = |e_i - e_j|$. In practice,  propensity scores are unknown and we use  logistic regression to estimate $e_i$s for the case studies in Sections \ref{sec:obesity} and \ref{sec:diet}.

\subsection{Propensity score matching}
We apply the propensity score matching algorithm for our case studies  in Sections \ref{sec:obesity} and \ref{sec:diet}. The simple weighted difference in means estimate for the ATE is given in the step 2 of the algorithm in Section \ref{sec:potentialoutcome}. 

We use matching with replacement to minimize the propensity score distance between the matched control individuals and the treatment individuals. This reduces bias, even if an individual in the control group is matched more than once.
As a comparison, matching without replacement  is sensitive to the order in which  individuals are matched.  This method may force  us to match individuals whose propensity scores are far apart, leading to an increase in bias. 
Further, 
We also use the single-nearest-neighbor matching, which selects a single individual in the control group whose propensity scores are closest to those of the treated individual. Single-nearest-neighbor matching can be extended to $k\geq 1$ nearest-neighbors. 

In addition to the simple weighted difference in means for estimating the treatment effect in the algorithm in Section \ref{sec:potentialoutcome}, one can also use a weighted regression, which takes account of the number of times a control is matched (see, e.g., Dehejia and Wahba \cite{dehejia2002}.)

\subsection{Model diagnosis}
The diagnosis of the quality of the resulting matched samples is an important step in using matching methods. 
In particular, we need to assess the covariate balance in terms of the similarity of
the empirical distributions of the full set of confounders
in the matched treated and control groups. 
Ideally, we want the  empirical distribution of $X_{T=1}$ in the treatment group  is the same as the  empirical distribution of $X_{T=0}$ in control treatment group . That is, the treatment is unrelated to the confounders. 

In our case studies  in Sections \ref{sec:obesity} and \ref{sec:diet}, we apply  the standardized difference in means as a balance measure: 
$(\bar{X}_{T=1} - \bar{X}_{T=0})/\sigma_{T=1}$,
where $\bar{X}_{T=1}$ and $\bar{X}_{T=0}$ are sample means of the treatment and control groups, and $\sigma_{T=1}$ is the sample standard deviation for the treatment group. 
We calculate the standardized difference in means for each covariate and use the $(\bar{X}_{T=1} - \bar{X}_{T=0})/\sigma_t<0.25$ as a criteria to check that the matching gives balanced samples (see, e.g., Rubin \cite{rubin2001}).


\section{Proofs}\label{app:proofs}
\subsection{Proof of Theorem \ref{prop:nonparaofdid}}
\begin{proof}

We claim that under the LCD-like treatment design (\ref{eqn:LCD-treatment}), it is not possible to obtain an accurate estimate of the treatment effect (\ref{eqn:nonparametricdid}) if the parametric functional forms of $f(\cdot)$ and $g(\cdot)$ in (\ref{eq:generalmodel}) are unknown. This claim is explained as follows. Under the null hypothesis that there is no trend in the control group $\{i|T_i(t=1)=0,T_i(t=0)=0\}$,  i.e.,
\begin{equation}
\label{eqn:nullhypothesis}
    \mathbb E_X[\mathbb E[Y_i(t=1)|T_i(t=1)=0,T_i(t=0)=0,X]] = \mathbb E_X[\mathbb E[Y_i(t=0)|T_i(t=1)=0,T_i(t=0)=0,X]],
\end{equation} then the difference-in-differences leads to an estimate to the following effect: 
\begin{equation}
\label{eqn:dideffect}
\begin{aligned}
   & \mathbb E_X[\mathbb E[Y_i(t=1)-Y_i(t=0)|T_i(t=1)=1,T_i(t=0)=0,X]]\\
& =    f(1) - f(0) + \mathbb E[g(X_{i}(t=1))]-\mathbb E[g(X_{i}(t=0))].
\end{aligned}
\end{equation}
Here $f(1) - f(0)$ corresponds to the treatment effect in (\ref{eqn:nonparametricdid}) and $\mathbb E[g(X_{i,1})]-\mathbb E[g(X_{i,0})]$ is the nuisance effect from the confounder. If the parametric functional forms of $f(\cdot)$ or $g(\cdot)$ is unknown, it is easy to show that for $g'(\cdot)\equiv 2g(\cdot)$, the treatment effect $f'(1)-f'(0)$ defined as follows satisfies (\ref{eqn:dideffect}):
\begin{equation*}
\begin{aligned}
   f'(1)-f'(0) &\equiv  \mathbb E_X[\mathbb E[Y_i(t=1)-Y_i(t=0)|T_i(t=1)=1,T_i(t=0)=0,X]] \\
   &\quad - \{\mathbb E[g'(X_{i}(t=1))]-\mathbb E[g'(X_{i}(t=0))]\}.
\end{aligned}
\end{equation*}
However, $ f'(1)-f'(0) \neq f(1) - f(0)$.
Hence, the treatment effect $f(1) - f(0)$ is unidentiable using difference-in-differences
under the two-point structure (\ref{eqn:LCD-treatment}). 
\end{proof}

\subsection{Multiple treatment versions}
\begin{figure}[!ht]
    \centering
    \includegraphics[width=0.5\textwidth]{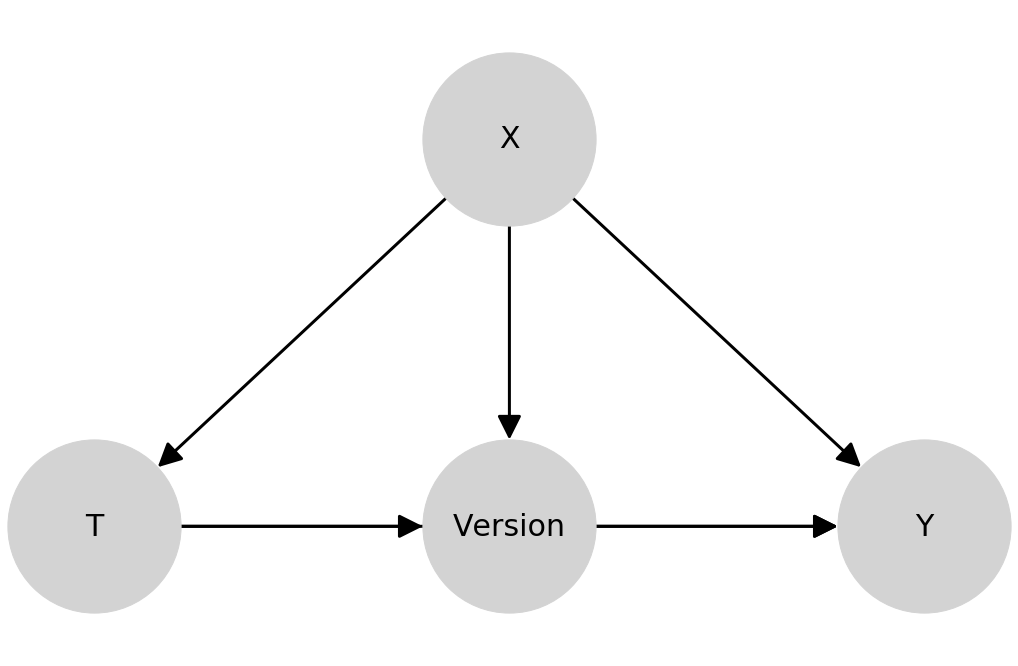}
    \caption{Causal graph illustrating relationship between treatment $T$, version of $T$, outcome $Y$, and $X$ consists of observed and unobserved confounders.}
    \label{fig:extra2}
\end{figure}

\begin{theorem}
\label{thm:propertyofirand}
Suppose that for each individual, there is a fixed version that would have been received, had the individual been given $T\in\{0,1\}$. Then
if Figure \ref{fig:extra2} is a causal graph, the average treatment effect is equivalent to
\begin{equation}
\label{eqn:avegoal}
    \text{ATE}^* \equiv \mathbb E_X[\mathbb E[Y|T=1,X]] - \mathbb E_X[\mathbb E[Y|T=0,X]].
\end{equation}
The I-Rand  Algorithm \ref{alg:i-rand} gives an unbiased estimate of $\text{ATE}^*$ in (\ref{eqn:avegoal}) if the estimator $\text{ATE}^{(m)}$ in (\ref{eqn:I-Rand}) is unbiased for ATE.
\end{theorem}
\begin{proof}
Since there is a fixed version of treatment that an individual would have been received if the individual has  been given $T\in\{0,1\}$, we have
\begin{equation*}
\begin{aligned}
    \mathbb E[Y(T)]  = \mathbb E_{X}[\mathbb E[Y(T)|X]] = \mathbb E_{X}[\mathbb E[Y(T)|T,X]].
\end{aligned}
\end{equation*}
where the last step is due to the fact that given Figure \ref{fig:extra2}, $T$ is ignorable relative to outcome $Y$, conditional on $X$ \cite{pearl2009causality}. Denote by $K^T(T)$ the counterfactual variable of which version of treatment that an individual would have been received if the individual has  been given $T\in\{0,1\}$. Then
\begin{equation*}
\begin{aligned}
 \mathbb E_{X}[\mathbb E[Y(T)|T,X]] & = \mathbb E_{X}[\mathbb E[Y(T,K^T(T))|T,X]] \\
 & = \mathbb E_{k^T,X}[\mathbb E[Y(T,k^T)|T,K^T(T)=k^T,X]] \\
  & = \mathbb E_{k^T,X}[\mathbb E[Y(T,k^T)|T,K^T=k^T,X]] \\
  & = \mathbb E_{k^T,X}[\mathbb E[Y|T,K^T=k^T,X]] \\
  & = \mathbb E_{X}[\mathbb E[Y|T,X]].
\end{aligned}
\end{equation*}
where the third step is by the assumption that there is a fixed version of treatment that an individual would have been received, and the third step is by the consistency for $Y$. Therefore, $\mathbb E[Y(T)] = \mathbb E_X[\mathbb E[Y|T,X]]$ and we obtain the desired the average treatment effect
\begin{equation*}
     \mathbb E_X[\mathbb E[Y|T=1,X]] - \mathbb E_X[\mathbb E[Y|T=0,X]].
\end{equation*}
Suppose that $X=(X_1,X_2)$, where $X_1$ consists of observed confounders and $X_2$ represents unobserved confounders. Then 
\begin{equation*}
\begin{aligned}
   &\mathbb E_X[\mathbb E[Y|T=1,X]] - \mathbb E_X[\mathbb E[Y|T=0,X]] = \sum_{x_2}\text{ATE}(X_2=x_2)\mathbb P(X_2=x_2).
\end{aligned}
\end{equation*}
where 
\begin{equation*}
\text{ATE}(X_2=x_2) = \mathbb E_{X_1}[\mathbb E[Y|T=1,X_1,X_2=x_2]]- \mathbb E_{X_1}[\mathbb E[Y|T=0,X_1,X_2=x_2]].
\end{equation*}
By the sampling strategy of the I-Rand estimator (\ref{eqn:I-Rand}) yields that $\mathbb P(X_2=x_2) = 2^{-N}$ and $\text{ATE}^{(m)}$ is a matching method estimator for $\text{ATE}(X_2=x_2)$. 
This completes the proof.
\end{proof}

\section{Variables Definition and Summary Statistics of Data Used in the Paper}\label{appendixA}

\begin{table}[!ht]
\let\center\empty
\let\endcenter\relax
\centering
\resizebox{0.7\width}{!}{\begin{tabular}{llrrrrrrrr}
\toprule
  &        &  count &     mean &     std &      min &      25\% &      50\% &      75\% &      max \\
LCD & Variable &        &          &         &          &          &          &          &          \\
\midrule
  & Gender &    256 &    0.590 &   0.493 &    0.000 &    0.000 &    1.000 &    1.000 &    1.000 \\
  & Age &    256 &   61.574 &  12.111 &   23.000 &   53.000 &   60.000 &   71.000 &   91.000 \\
  & Height &     75 &    1.706 &   0.092 &    1.473 &    1.625 &    1.720 &    1.770 &    1.900 \\
  & Weight &    251 &   96.160 &  18.621 &   55.300 &   83.700 &   95.000 &  107.000 &  159.000 \\
  & BMI &     66 &   33.887 &   6.071 &   21.660 &   29.890 &   33.495 &   36.980 &   57.100 \\
0  & T2D &    256 &    1.281 &   0.811 &    0.000 &    1.000 &    2.000 &    2.000 &    2.000 \\
  & HbA1c/   mmol/mol &    202 &   61.376 &  20.652 &   37.000 &   45.000 &   54.500 &   71.000 &  135.000 \\
  & TBC &    176 &    5.314 &   1.302 &    2.500 &    4.385 &    5.200 &    6.225 &    9.300 \\
 & HDL &    195 &    1.280 &   0.421 &    0.600 &    1.000 &    1.200 &    1.450 &    3.500 \\
  & SBP &    171 &  143.503 &  15.476 &  114.000 &  132.000 &  142.000 &  152.000 &  223.000 \\
  \midrule
  & Gender &    256 &    0.590 &   0.493 &    0.000 &    0.000 &    1.000 &    1.000 &    1.000 \\
  & Age &    256 &   63.424 &  12.387 &   23.167 &   54.750 &   62.750 &   73.500 &   91.500 \\
  & Height &     75 &    1.706 &   0.092 &    1.473 &    1.625 &    1.720 &    1.770 &    1.900 \\
  & Weight &    251 &   87.070 &  17.352 &   51.000 &   75.000 &   84.400 &   97.100 &  140.000 \\
  & BMI &     65 &   30.356 &   5.923 &   19.240 &   27.040 &   29.270 &   32.470 &   53.620 \\
1  & T2D &    256 &    0.719 &   0.867 &    0.000 &    0.000 &    0.000 &    2.000 &    2.000 \\
  & HbA1c/   mmol/mol &    201 &   45.925 &   9.319 &   32.000 &   40.000 &   43.000 &   50.000 &   84.000 \\
  & TBC &    174 &    4.892 &   1.247 &    2.400 &    4.025 &    4.700 &    5.700 &    8.800 \\
 & HDL &    189 &    1.413 &   0.542 &    0.700 &    1.090 &    1.340 &    1.610 &    4.900 \\
  & SBP &    170 &  132.100 &  11.021 &  108.000 &  125.000 &  132.000 &  139.500 &  170.000 \\
  & months &    256 &   22.199 &  17.456 &    1.000 &    8.000 &   19.000 &   32.000 &   84.000 \\
\bottomrule
\end{tabular}
}
\caption{Summary statistics of variables collected in the study. LCD=0 corresponds to data collected at the first visit and LCD=1 for data collected at the second visit.}
\label{tab:table_variables}
\end{table}

\noindent \textbf{Gender}:  a binary variable with ``female”$ =0$ and ``male” $=1$.\\
\textbf{Age}: the age of the participants at their visit.\\
\noindent \textbf{BMI}: the body mass index of the participants. Here BMI is defined as the ratio of the weight squared height. We note that although recent studies on nutrition suggest
that different obesity metrics can lead to different relationships between obesity
to CVD risk,
the consensus is that compared to BMI measures the more refined modalities (e.g., waist circumference, waist-to-hip ratio, waist-to-height ratio) do not add significantly to the BMI assessment from a clinical perspective  \cite{gelber2008}. \\
\textbf{T2D}: a three-states variable to inform of the type-2 diabetes status; 0 for non-diabetic, 1 for pre-diabetic and 2 for diabetic. \\
\textbf{HbA1c}: the glycated haemoglobin of the participants. It develops when haemoglobi, a protein within red blood cells that carries oxygen throughout the body, joins with glucose in the blood, becoming `glycated'. This measure allows to determine the T2D status. \\
\textbf{LCD}: a binary variable which equals to $1$ only if the participant is suggested to follow a low-carbohydrate diet.\\
\textbf{TBC}: the total blood cholesterol level of the participants. It is a measurement of certain elements in the blood, including the amount of high- and low-density lipoprotein cholesterol (HDL and LDL) in a person's blood.
\\
\textbf{HDL}: the high-density lipoprotein cholesterol of the participants. The HDL is the well-behaved "good cholesterol." This friendly scavenger cruises the bloodstream. As it does, it removes harmful ``bad" cholesterol from where it doesn't belong. A high HDL level reduces the risk for heart disease.\\
\textbf{SBP}: the systolic blood pressure of the participants. The SBP indicates how much pressure the blood is exerting against your artery walls when the heart beats. It is one of the CVD risk factors used to calculate the Reynolds risk score.\\
\textbf{Months}: the number of months between the two visits of participants to the clinic (end date – start date).\\

\section{Supplementary Numerical Results}
\label{sec:suppnum}

\subsection{Treatment effect of LCD on T2D}
\label{sec:lcdt2d}

\begin{figure}[!ht]
    \centering
    \includegraphics[width=0.5\textwidth]{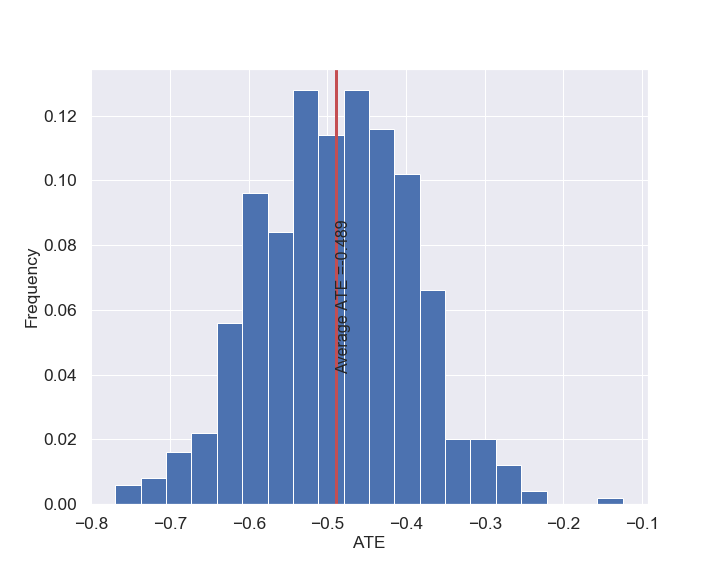}
    \caption{Distribution of the  ATE of the LCD on T2D in Figure \ref{fig:dag_lcd_t2d}. The results are based on 500 subsamples.}
    \label{fig:lcdt2d_subsampling}
\end{figure}

\begin{figure}[!ht]
    \centering
    \includegraphics[width=0.5\textwidth]{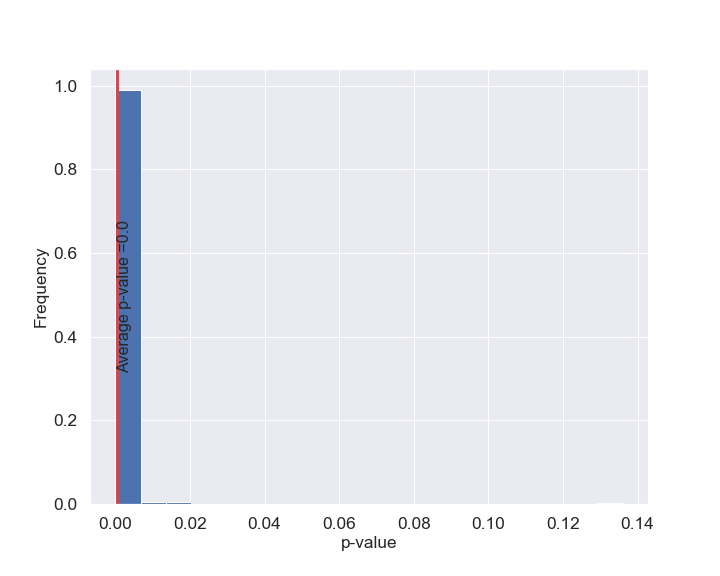}
    \caption{Distribution of p-values of the ATE of the LCD on T2D in Figure \ref{fig:dag_lcd_t2d}. The results are based on 500 subsamples.}
    \label{fig:lcdt2d_pvalue}
\end{figure}

We provide details on assessing the significance of the reduction of the risk of T2D due to the LCD using the  I-Rand algorithm, where the causal diagram is shown in Figure \ref{fig:dag_lcd_t2d}. We show
the distribution of ATEs for the subsampling step in Figure \ref{fig:lcdt2d_subsampling} and the distribution of the p-value from the permutation test under the null hypothesis of no causal effect (ATE = 0) in Figure \ref{fig:lcdt2d_pvalue}. 
The distributions confirm the consistency of these results across the subsamples.

\subsection{Mediation analysis for the effect of LCD on CVD}
\label{sec:lcdcvd}

\begin{figure}[!ht]
    \centering
    \includegraphics[width=0.5\textwidth]{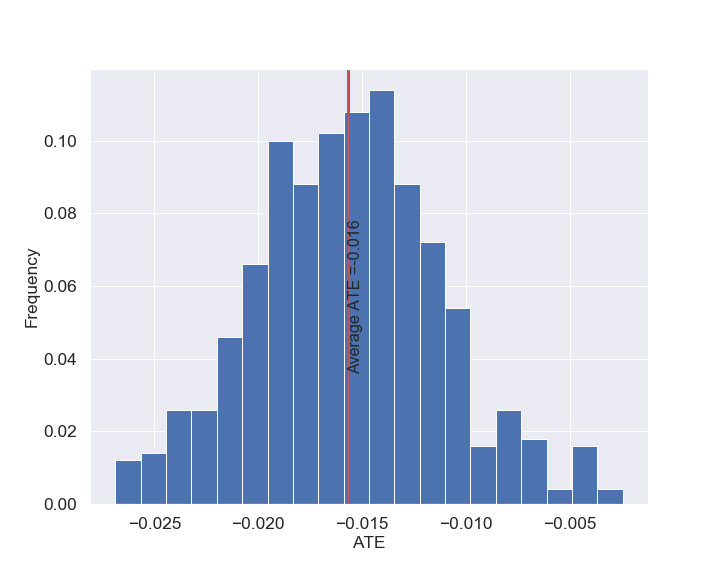}
    \caption{Distribution of total effect of the LCD on the Reynold risk score in Figure \ref{fig:dag_lcd_cvd2}.  The results are from 500 subsamples.}
    \label{fig:lcdcvd_subsampling1}
\end{figure}

\begin{figure}[!ht]
    \centering
    \includegraphics[width=0.5\textwidth]{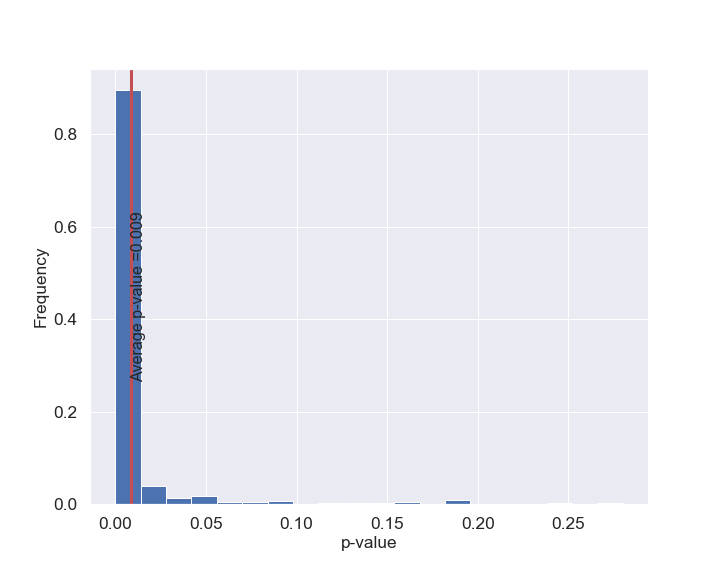}
    \caption{Distribution of p-values the total effect of the LCD on the Reynolds risk score in Figure \ref{fig:dag_lcd_cvd2}.  The results are based on 500 subsamples.}
    \label{fig:lcdcvd_pvalue1}
\end{figure}

\begin{figure}[!ht]
    \centering
    \includegraphics[width=0.5\textwidth]{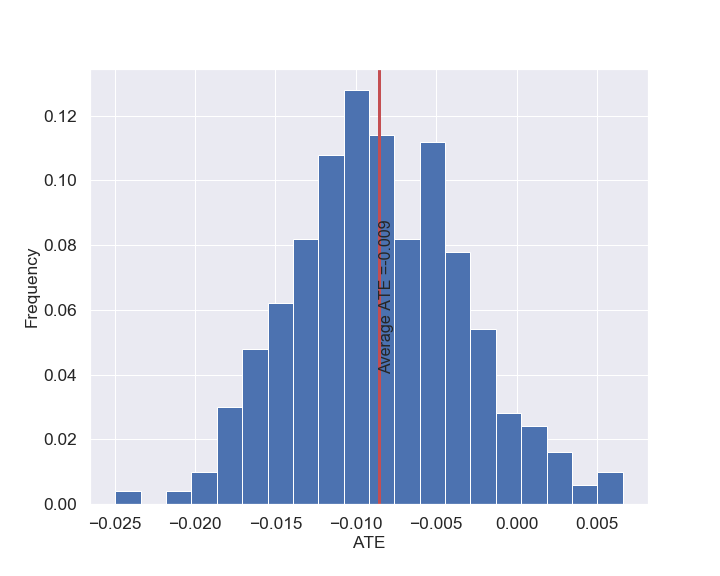}
    \caption{Distribution of the direct effect of the LCD on the Reynolds risk score in Figure \ref{fig:dag_lcd_cvd2}.  The results are based on 500 subsamples.}
    \label{fig:lcdcvd_subsampling2}
\end{figure}

\begin{figure}[!ht]
    \centering
    \includegraphics[width=0.5\textwidth]{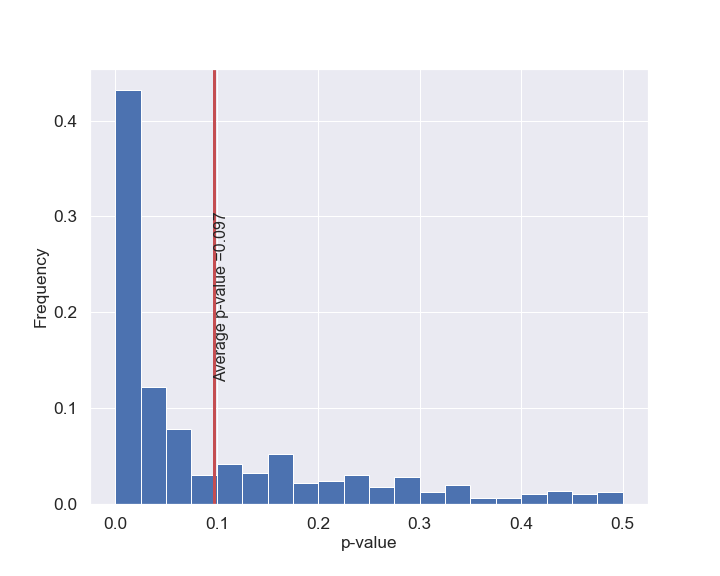}
    \caption{Distribution of  p-values of the direct effect of the LCD on the Reynold risk score in Figure \ref{fig:dag_lcd_cvd2}.  The results are based on 500 subsamples.}
    \label{fig:lcdcvd_pvalue2}
\end{figure}

We  provide  details  on  assessing  the  significance  of  reduction in Reynolds risk score due to the  low-carbohydrate using the  I-Rand algorithm. 
The causal diagrams are shown in Figure \ref{fig:dag_lcd_cvd2} for the direct and indirect effects. 
We show the distribution of ATE for the subsampling step in Figure \ref{fig:lcdcvd_subsampling1} and the distribution of the p-values from the permutation test under the null hypothesis of no causal effect (ATE = 0) in Figure \ref{fig:lcdcvd_pvalue1}, for the total effect (sum of direct and indirect effect); and correspondingly, Figures \ref{fig:lcdcvd_subsampling2} and \ref{fig:lcdcvd_pvalue2}, for the direct effect.  The distributions confirm the consistency of these results across the subsamples.

\subsection{Causal effect of obesity on T2D}

\begin{figure}[!ht]
    \centering
    \includegraphics[width=0.6\textwidth]{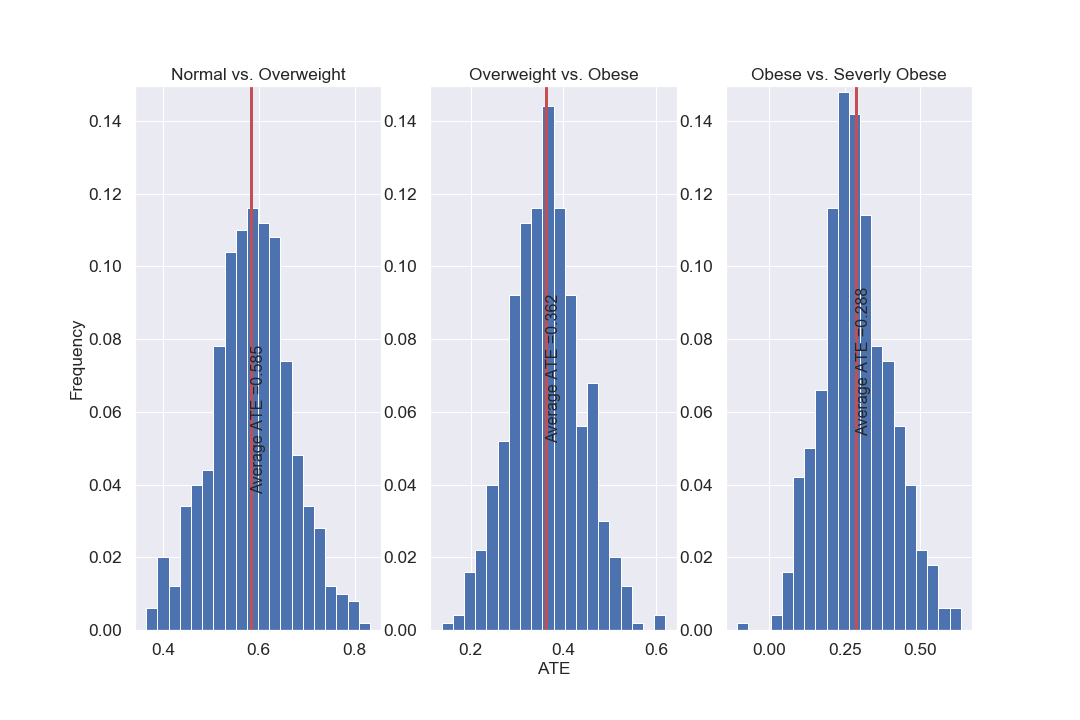}
    \caption{Distributions of ATE of  obesity on T2D in Figure \ref{fig:bmit2d}. The results are based on 500 subsamples with different BMI splits.}
    \label{fig:bmit2d_subsampling}
\end{figure}

\begin{figure}[!ht]
    \centering
    \includegraphics[width=0.6\textwidth]{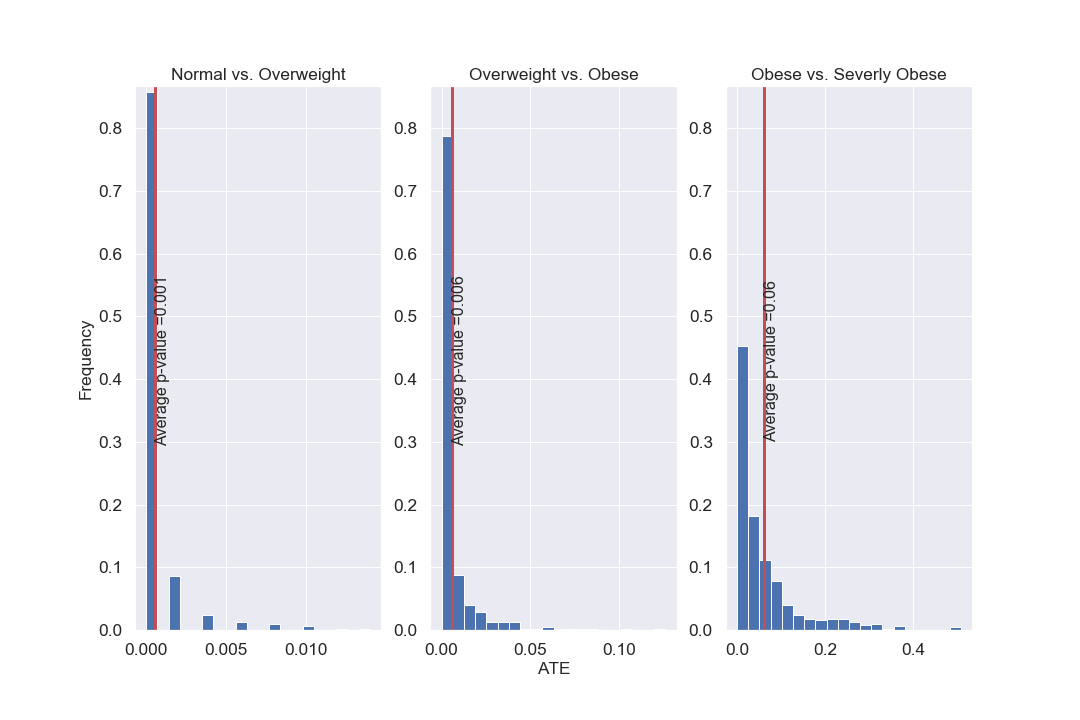}
    \caption{Distributions of p-values of the ATE of obesity on T2D in Figure \ref{fig:bmit2d}. The results are based on 500 subsamples with different BMI splits.}
    \label{fig:bmit2d_pvalue}
\end{figure}

We provide additional details on testing the significance of obesity as a cause of T2D, where the causal diagram is shown in Figure \ref{fig:bmit2d}. In particular, we show
the distribution of ATE for the subsampling step in Figure \ref{fig:bmit2d_subsampling} and the distribution of p-values from the permutation test under the null hypothesis of no causal effect (ATE = 0) in Figure \ref{fig:bmit2d_pvalue}, using the I-Rand algorithm. 
The distributions confirm the consistency of these results across the subsamples. 

\subsection{Mediation analysis for the effect of obesity on CVD}
\label{sec:obsitycvd}

\begin{figure}[!ht]
    \centering
    \includegraphics[width=0.6\textwidth]{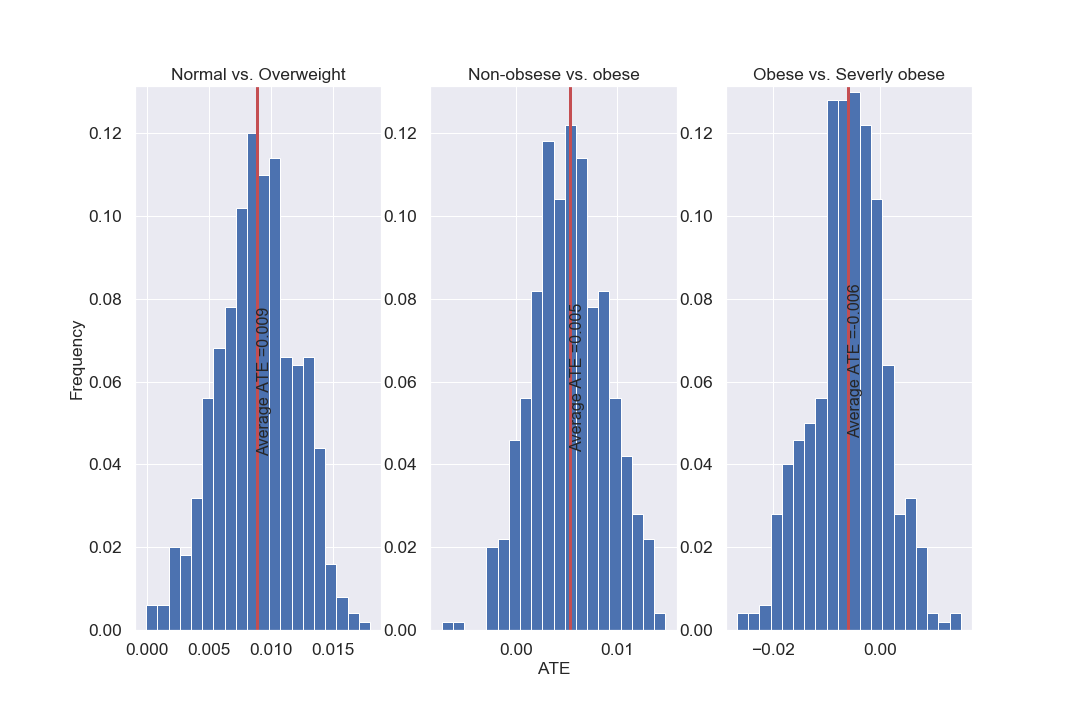}
    \caption{Distributions of the total effect of  obesity on the systolic blood pressure in Figure \ref{fig:dag_bmi_cvd2}. The results are based on 500 subsamples with different BMI splits.}
    \label{fig:bmicvd_subsampling1}
\end{figure}

\begin{figure}[!ht]
    \centering
    \includegraphics[width=0.6\textwidth]{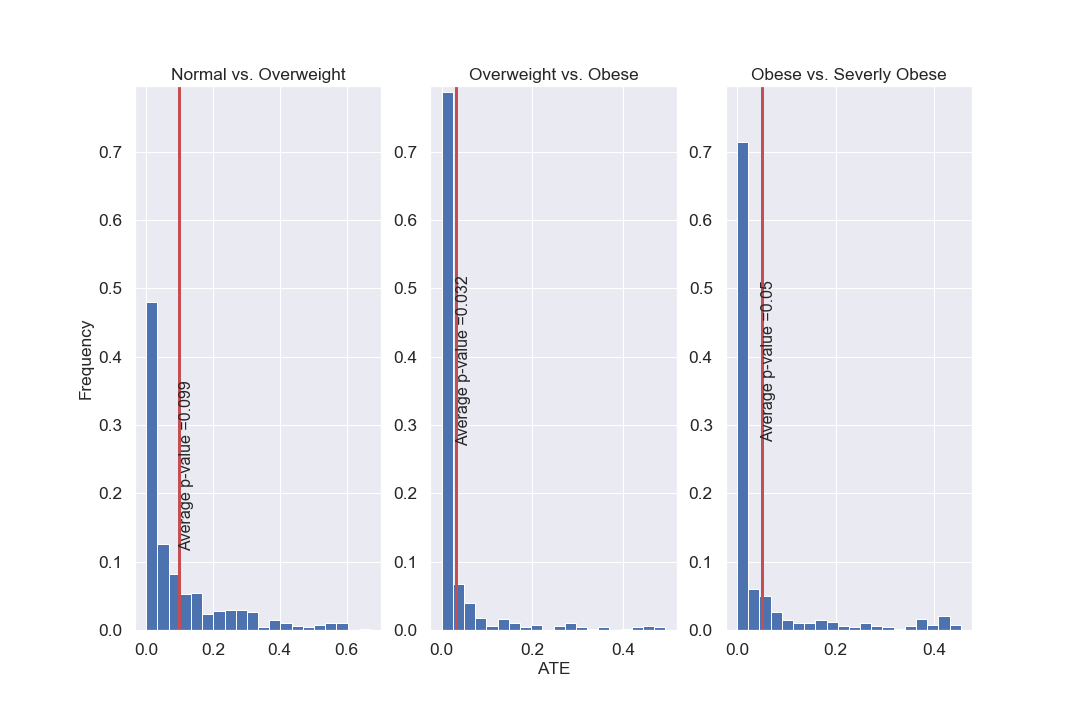}
    \caption{Distributions of p-values of the total effect of obesity on the systolic blood pressure in Figure \ref{fig:dag_bmi_cvd2}. The results are based on 500 subsamples with different BMI splits.}
    \label{fig:bmicvd_pvalue1}
\end{figure}

\begin{figure}[!ht]
    \centering
    \includegraphics[width=0.6\textwidth]{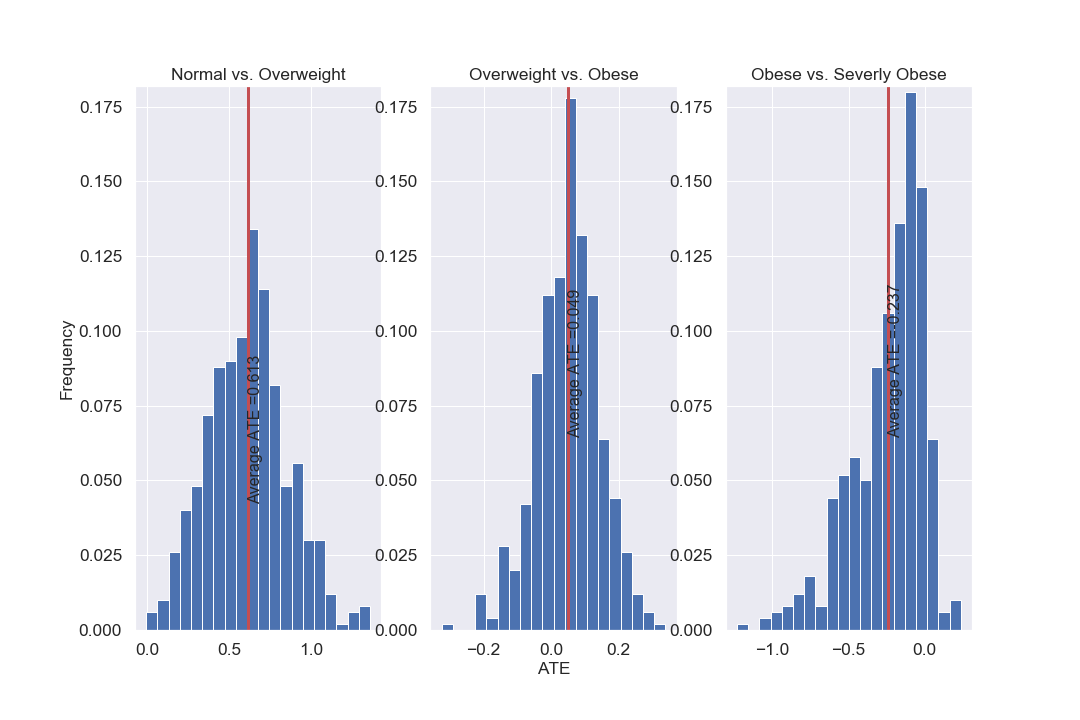}
    \caption{Distributions of  the direct  effect  of obesity on the systolic blood pressure in Figure \ref{fig:dag_bmi_cvd2}. The results are based on 500 subsamples with different BMI splits.}
    \label{fig:bmicvd_subsampling2}
\end{figure}

\begin{figure}[!ht]
    \centering
    \includegraphics[width=0.6\textwidth]{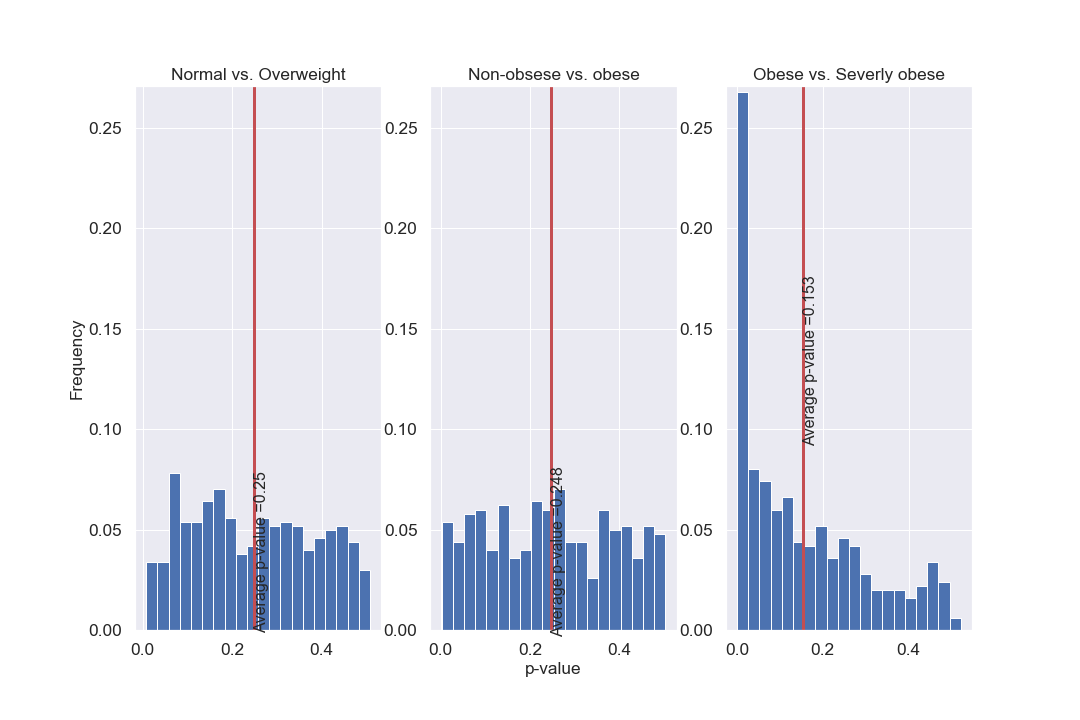}
    \caption{Distribution of p-values of the  direct effect of  obesity on the systolic blood pressure in Figure \ref{fig:dag_bmi_cvd2}. The results are based on 500 subsamples with different BMI splits.}
    \label{fig:bmicvd_pvalue2}
\end{figure}

We provide results on testing the significance of. the effect of obesity on high systolic blood pressure, according to the proposed I-Rand algorithm.  The causal diagrams are shown in Figure \ref{fig:dag_bmi_cvd2} for the direct and indirect effects. In particular, we show the distribution of ATE for the subsampling step in Figure \ref{fig:bmicvd_subsampling1} and the distribution of the p-value from the permutation test under the null hypothesis of no causal effect (ATE = 0) in Figure \ref{fig:bmicvd_pvalue1},  for the total effect; and correspondingly, Figures \ref{fig:bmicvd_subsampling2} and \ref{fig:bmicvd_pvalue2}, for the direct effect. 
The distributions confirm the causal effect of the obesity to CVD consistently across the subsamples. 

\end{document}